\documentclass[letterpaper, 10 pt, conference]{IEEEtran}
\IEEEoverridecommandlockouts  

\usepackage[utf8]{inputenc} 
\usepackage[T1]{fontenc}    
\usepackage{url}            
\usepackage{amsfonts}       
\usepackage{amsmath} 
\usepackage{amssymb}  
\usepackage{amsthm}
\usepackage{mathrsfs}
\usepackage[ruled]{algorithm2e}
\usepackage{bm}
\usepackage{stfloats}
\usepackage{svg}
\usepackage{graphicx}
\usepackage{threeparttable}
\usepackage{nicefrac}       
\usepackage{microtype}      
\usepackage[colorlinks=true, allcolors=black]{hyperref}
\usepackage[normalem]{ulem}
\usepackage{caption}
\usepackage{subcaption}

\newcommand{\mS}{\mathcal{S}}
\newcommand{\mX}{\mathcal{X}}
\newcommand{\mU}{\mathcal{U}}
\newcommand{\mR}{\mathcal{R}}
\newcommand{\R}{\mathbb{R}}
\newcommand{\mbE}{\mathbb{E}}
\newcommand{\mbP}{\mathbb{P}}

\newcommand{\mO}{\mathcal{O}}

\newcommand{\tX}{\tilde{X}}
\newcommand{\tx}{\tilde{x}}
\newcommand{\ty}{\tilde{y}}

\newcommand{\salf}[1]{\Phi_{n,\lambda}(#1)}

\newcommand{\defeq}{:=}

\newcommand{\cov}[1]{\text{cov}(#1)}

\newcommand{\innerp}[1]{\langle #1 \rangle}
\newcommand{\expect}[1]{\mathbb{E}\left( #1 \right)}
\newcommand{\expectw}[2]{\mathbb{E}_{#1}\left( #2 \right)}
\newcommand{\prob}[1]{\mathbb{P}\left( #1 \right)}
\newcommand{\ball}[1]{\mathcal{B}^{n}\left( #1 \right)}

\newtheorem{definition}{Definition}[section]
\newtheorem{lemma}{Lemma}[section]
\newtheorem{proposition}{Proposition}
\newtheorem{theorem}{Theorem}
\newtheorem{assumption}{Assumption}
\newtheorem{remark}{Remark}[section]
\newtheorem{problem}{Problem}

\usepackage{pdfpages}
\usepackage{stmaryrd}

\newcommand{\OF}{\mathsf{F}}

\newcommand{\tr}{\mathrm{tr}}

 \newcommand{\setdef}[2]{\{#1
	\; | \; #2\}}

\newcommand\oprocendsymbol{\hbox{$\triangle$}}
\newcommand\oprocend{\relax\ifmmode\else\unskip\hfill\fi\oprocendsymbol}

\let\labelindent\relax
\usepackage{enumitem} \renewcommand{\theenumi}{(\roman{enumi})}

\DeclareSymbolFont{bbold}{U}{bbold}{m}{n}
\DeclareSymbolFontAlphabet{\mathbbold}{bbold}

\newcommand{\real}{\mathbb{R}}

\newcommand{\seminorm}[1]{{\left\vert\kern-0.25ex\left\vert\kern-0.25ex\left\vert #1
		\right\vert\kern-0.25ex\right\vert\kern-0.25ex\right\vert}}

\newcommand{\semimeasure}[1]{\mu_{\seminorm{\cdot}}\kern-0.5ex\left(#1\right)}

\renewcommand{\top}{\mathsf{T}} 

\newcommand{\suchthat}{\;\ifnum\currentgrouptype=16 \middle\fi|\;}
\newcommand{\scirc}{\raise1pt\hbox{$\,\scriptstyle\circ\,$}}

\title{\LARGE \bf
Probabilistic Reachability Analysis of Stochastic Control Systems
}
\author{Saber Jafarpour$^{1*}$, Zishun Liu$^{2*}$ and Yongxin Chen$^{2}$
\thanks{The first two authors contribute equally to this work}
\thanks{$^{1}$ Saber Jafarpour is with University of Colorado Boulder, Boulder, CO 80309
        {\tt\small Saber.Jafarpour@colorado.edu}}%
\thanks{$^{2}$Zishun Liu and Yongxin Chen are with Georgia Institute of Technology, Atlanta, GA 30332 
        {\tt\small \{zliu910\}\{yongchen\}@gatech.edu}}%
}

\begin{document}
\maketitle
\thispagestyle{empty}
\pagestyle{empty}

\begin{abstract}
We address the reachability problem for continuous-time stochastic dynamic systems. 
Our objective is to present a unified framework that characterizes the reachable set of a dynamic system in the presence of both stochastic disturbances and deterministic inputs. 
To achieve this, we devise a strategy that effectively decouples the effects of deterministic inputs and stochastic disturbances on the reachable sets of the system. 
For the deterministic part, many existing methods can capture the deterministic reachability. 
As for the stochastic disturbances, we introduce a novel technique that probabilistically bounds the difference between a stochastic trajectory and its deterministic counterpart.
The key to our approach is introducing a novel energy function termed the Averaged Moment Generating Function that yields a high probability bound for this difference. 
This bound is tight and exact for linear stochastic dynamics and applicable to a large class of nonlinear stochastic dynamics. 
By combining our innovative technique with existing methods for deterministic reachability analysis, we can compute estimations of reachable sets that surpass those obtained with current approaches for stochastic reachability analysis. 
We validate the effectiveness of our framework through various numerical experiments. Beyond its immediate applications in reachability analysis, our methodology is poised to have profound implications in the broader analysis and control of stochastic systems. It opens avenues for enhanced understanding and manipulation of complex stochastic dynamics, presenting opportunities for advancements in related fields.
\end{abstract}

\begin{IEEEkeywords}
Reachability analysis, stochastic dynamic systems, stochastic control
\end{IEEEkeywords}

\section{Introduction}
\label{sec:introduction}

Reachability analysis is an important topic in systems and control theory that focuses on analyzing whether the trajectory of a system will reach a certain set within a time horizon starting from a given set of initial conditions and possibly subject to inputs or disturbances. It is essential in many applications including autonomous vehicles, aerospace systems, robotics, etc. For instance, in safety-critical applications where the system should be kept outside an unsafe region of the state space, reachability analysis is a key machinery to verify and design the control input to avoid the unsafe region. 

Reachability analysis of dynamical systems is a fundamentally challenging task. For general dynamical systems, obtaining exact or close approximations of their reachable sets is only possible when the state dimension is low and generally demands substantial computational resources. 
However, there is a rapidly growing need for fast reachability analysis methods in various control applications.
This motivates the need for rigorous methods that can efficiently upper bound the reachable sets of dynamical systems. 

For deterministic systems with bounded inputs or disturbances, many methods have been proposed to over-approximate the reachable set. 
Several representative methods include Hamilton-Jacobi reachability that poses reachability as a game between two players~\cite{SB-MC-SH-CJT:17,IM:07}, contraction-based reachability that estimates the propagation of reachable sets using contraction rate of the system~\cite{JM-MA:15,FB:22-CTDS}, and Interval-based reachability that over-approximates reachable sets by leveraging techniques from interval analysis and monotone system theory~\cite{scott2013bounds,PJM-AD-MA:19,SC:20,SJ-AH-SC:23}. Other methods such as simulation-based reachability \cite{chuchu2017simulation,huang2012computing} are also popular in a wide range of studies. %

In this work, we are interested in reachability analysis for stochastic systems. In many real-world applications, systems are subject to unbounded and stochastic disturbances and are better modeled by stochastic dynamics. 
Despite the efficiency of the aforementioned deterministic reachability methods in the presence of bounded disturbances, they cannot be applied directly to systems subject to unbounded stochastic disturbances. 
For systems with stochastic disturbances, considering all possible disturbance scenarios will often result in unbounded reachable sets due to the unboundedness of stochastic noise. Moreover, this approach also ignores the statistical properties of the noise, leading to overly conservative results~\cite{cosner2024bounding}.
To better capture the effects of stochastic disturbances, reachability analysis in stochastic systems focuses on the \textit{probabilistic reachable set}, which refers to the set that any possible trajectory starting from an initial set can reach with high probability (e.g., 99.9\%). 

There have been several attempts to approximate probabilistic reachable sets of stochastic systems and they can be divided into two categories. 
The first category is the optimization-based approaches that use Hamilton-Jacobi equations and dynamic programming~\cite{HMS-NT:02,AA-MP-JL-SS:08,SS-JL:10,PM-DC-JL:16}. 
However, these approaches are usually computational heavy, rendering them impractical for large-scale systems. 
The second category is simulation-based approaches which provide guarantees for reachability using trajectory samples~\cite{KL-MO-RSE:13,HS-APV-BA-MO:19,NH-XQ-LL-JVD:23}. One drawback of these methods is that the amount of samples needed to obtain reasonable bounds on reachable sets can grow exponentially.
Another tangentially related line of research is on the stochastic Lyapunov function \cite{black2024risk} or barrier function \cite{HE-MF-XL-AP-SP:06,SP-AJ-GJP:07, santoyo2021barrier,MA-AL-MZ:22} for measuring the probability of a trajectory staying in a safe set. In these works, the goal is not to find the probabilistic reachable set but to verify whether a given safe set is in the probabilistic reachable set.

In this work, we establish a unified framework for computing the probabilistic reachable sets of nonlinear stochastic systems subject to both deterministic inputs and stochastic disturbances. 
Our method is both theoretically optimal and effective in practice. Theoretically, under standard assumptions, our method yields tight approximations of probabilistic reachable sets that cannot be improved further without additional assumptions. Implementation-wise, our approximations of probabilistic reachable sets can be computed efficiently and are scalable to high-dimensional systems. 

Our framework is built upon a novel separation strategy, which decouples the effects of deterministic inputs and stochastic uncertainty on reachability analysis of the stochastic system (Proposition 1). 
The effects of stochastic uncertainty on the probabilistic reachable set can be represented using stochastic deviation, which refers to the distance between a stochastic trajectory and its associated deterministic trajectory.
By developing a novel energy function termed the Averaged Moment Generating Function (AMGF), we provide a probabilistic bound on the stochastic deviation of general stochastic continuous trajectories (Theorem \ref{thm: sto vib}). 
Our bound has a dependence $\mO(\sqrt{\log (1/\delta)})$ on the probability level $1-\delta$, significantly better than existing techniques which result in a bound of the order $\mO(\sqrt{1/\delta})$. Moreover, our bound coincides with that for linear stochastic systems under the same assumptions and cannot be improved further.
The effects of deterministic input on the probabilistic reachable set can be captured using deterministic reachable sets of the associated deterministic system, i.e., the system obtained by removing the stochastic noise. 
Consequently, our separation strategy enables a decomposition of the probabilistic reachable set into a deterministic reachable set capturing the deterministic input and a tight robustness buffer around it against the stochastic uncertainty (Theorem \ref{thm: PRS}). As such, analyzing the reachability of the associated deterministic system is all we need to obtain a good probabilistic reachable set. 
This is a paradigm shift and brings tremendous flexibility to the reachability analysis of stochastic systems as any deterministic reachability framework can be incorporated. 
In particular, we combine our framework with two computationally efficient deterministic reachability approaches namely contraction-based reachability and interval-based reachability to obtain probabilistic reachable sets for stochastic systems. 

Finally, our tight probabilistic bound of stochastic deviation is poised to have profound implications in the broader analysis and control of stochastic systems beyond its immediate applications in reachability analysis. To the best of our knowledge, this bound is the first non-conservative result that can quantitatively describe the behavior of a nonlinear stochastic system under standard assumptions.
The bound is of independent interests and can potentially impact many other areas such as estimation, uncertainty quantification, finance, machine learning, statistics, etc. It opens avenues for enhanced understanding and manipulation of complex stochastic dynamics, presenting opportunities for advancements in related fields.

The rest of the paper is organized as follows. In Section \ref{sec: bg} we briefly review reachability analysis for deterministic systems. In Section \ref{sec: formulation} we introduce and formulate the probabilistic reachability problem and present our overall strategy. The discussion of an existing method is given in Section \ref{sec: limit}. Section \ref{sec:propag} contains the main technical contribution of this paper where we introduce a novel energy function termed the Averaged Moment Generating Function to bound the deviation of stochastic trajectories from their deterministic counterpart with high-probability. This high-probability bound of stochastic deviation is combined with deterministic reachability analysis in Section \ref{sec: PRS} to approximate the probabilistic reachable set of stochastic systems. This is followed by case studies in Section \ref{sec: app} and numerical experiments in Section \ref{sec: simulations}.

\section{Preliminaries} \label{sec: bg}
In this section, we briefly review reachability analysis for deterministic dynamics and related concepts. 

\subsection{Notations}
\textit{Vectors and matrices.} Given a vector $x\in\R^n$, $\|x\|$ denotes its Euclidean norm ($\ell_2$ norm) and $\|x\|_P=\sqrt{x^{\top}Px}$ with some positive definite matrix $P$. Given a matrix $A\in\R^{m\times n}$, $\|A\|$ denotes its spectral norm and $\|A\|_P$ denotes its weighted spectral norm with respect to some positive definite matrix $P$. For two matrices $A,B\in\R^{n\times n}$, $A\preceq B$ if $B-A$ is positive semi-definite. If $A\in\R^{n\times n}$ is a positive definite matrix, we denote its square root by $A^{\frac{1}{2}}$, i.e., $A^{\frac{1}{2}}$ is the unique matrix such that $A^{\frac{1}{2}}(A^{\frac{1}{2}})^{\top} = (A^{\frac{1}{2}})^{\top}A^{\frac{1}{2}} =A$. Besides, we use $\innerp{\cdot,\cdot}$ to denote standard inner product, $0$ to denote all-zero vectors and matrices, and $I_n$ to denote $n$-dimensional identity matrix. 

\textit{Set and Functions.} We use $\ball{r,y}$ to denote the ball $\{x\in\R^n: \|x-y\|\leq r\}$ and $\mS^{n-1}$ to denote the unit sphere $\{x\in\R^n:\|x\|=1\}$. For two sets $A,B$, their Minkowski sum is defined as $A\oplus B = \{x+y: x\in A,~ y\in B\}$. 
Given a set $\mathcal{X}\subseteq \real^n$ and a matrix $T\in \real^{n\times n}$, we define $T\mathcal{X}=\setdef{Tx}{x\in \mathcal{X}}$.
Given a continuously differentiable vector-valued function $f:\real^n\to \real^m$, we denote the Jacobian of $f$ at $x$ by $D_xf(x)$. For a twice-differentiable scalar-valued function $f:\real^n\to \R$, its gradient at $x$ is $\nabla f(x)$ and the Hessian matrix is denoted as $\nabla^2f(x)$. 

Throughout the paper, we use $\mbE$ to denote expectation and $\mbP$ to denote probability. For a set $S$, $X\sim S$ means $X$ is a random variable drawn uniformly from $S$.

\subsection{Reachable Set of Deterministic Dynamics}
Computing the reachable sets is a fundamental problem in dynamical systems and control theory. 
Consider the continuous-time deterministic system 
\begin{equation}\label{eq:deterministic}
    \dot{x}_t=f(x_t,u_t,t),
\end{equation}
where $x_t\in \R^n$ is the state at time $t$, $u_t\in \R^{p}$ is the input at time $t$, and $f: \R^n\times\R^p\times\real_{\ge 0}\to\R^n$ is a parameterized vector field. 
Depending on the applications, $u_t$ can be a  control action or a disturbance. 
The reachable set of a deterministic system is the set of all states that the system can reach, starting from an initial configuration, under all possible inputs within a specified time period~\cite{XC-SS:22}. 
\begin{definition}[DRS]\label{def: DRS}
    Consider the system~\eqref{eq:deterministic} with initial set $\mathcal{X}_0\subseteq \real^n$ and input set $\mathcal{U}\subseteq \real^p$. The \textit{deterministic reachable set} (DRS) of~\eqref{eq:deterministic} at time $t$ starting from $\mathcal{X}_0$ with inputs in $\mathcal{U}$ is
\begin{align}\label{eq:reach}
\mathcal{R}_t = \left\{ x_t \middle|
    \begin{aligned}
    &\tau\mapsto x_\tau \mbox{ is a trajectory of~\eqref{eq:deterministic}}\\ &\mbox{ with } x_0\in \mathcal{X}_0 \mbox{ and } u_\tau: \real_{\ge 0}\to \mathcal{U}
    \end{aligned}
    \right\}
\end{align} 
\end{definition}

In general, computing the exact DRS of a dynamic system is computationally intractable~\cite{CM:90}. Therefore, most methods in reachability analysis focus on providing over-approximation of DRS~\cite{XC-SS:22}. A set $\overline{\mathcal{R}}_t\subseteq \real^n$ is an over-approximation of the DRS~\eqref{eq:reach} if, for every $t\ge 0$, 
\begin{align*}
    \mathcal{R}_t\subseteq \overline{\mathcal{R}}_t.
\end{align*}
In Section~\ref{sec: app}, we revisit two approaches to compute $\overline{\mathcal{R}}_t$: contraction-based reachability and interval-based reachability.

\subsection{Matrix Measure and Contraction Theory}

A key tool in studying reachable sets of system \eqref{eq:deterministic} is the \textit{matrix measure}~\cite{CAD-HH:72,GS:06} defined as follows.
\begin{definition}[Matrix Measure]\label{def:matrix}
    Given a matrix $A\in\R^{n\times n}$, its matrix measure with respect to $\|\cdot\|$, denoted by $\mu(A)$, is defined as
    \begin{align*}
        \mu(A)=\lim_{\epsilon\to 0^{+}}\frac{\|I_n + \epsilon A\|-1}{\epsilon}.
    \end{align*}
\end{definition}
Intuitively, $\mu(A)$ can be considered as the one-sided derivative of the norm $\|\cdot\|$ at $I_n$ in the direction of $A$. 
Although matrix measure can be defined with respect to any norm, in this paper we focus on the spectral norm. In this case, the matrix measure has a closed-form expression $\mu(A) = \tfrac{1}{2}\lambda_{\max}(A+A^{\top})$. 

For the system \eqref{eq:deterministic}, the evolution of the distance of two arbitrary trajectories can be measured using $\mu(D_xf(x,u,t))$. The following lemma provides a variational characterization of $\mu(D_xf(x,u,t))$~\cite{AD-SJ-FB:20o}. 

\begin{lemma}\label{lemma: matrix measure}
    Given a deterministic system \eqref{eq:deterministic},
    for every $t\ge 0$, the following statement are equivalent 
    \begin{enumerate}
        \item\label{cond1} $\mu(D_xf(x,u,t)) \leq c_t$ for all $(x,u) \in \R^n\times \mU$. 
        \item\label{cond2} $(x-y)^{\top}(f(x,u,t)-f(y,u,t))\leq c_t\|x-y\|^2$, for all $(x,y,u) \in \R^n\times \real^n\times \mU$. 
    \end{enumerate}
\end{lemma}
A classical result in contraction theory states that if condition~\ref{cond1} holds then the distance between trajectories of the system~\eqref{eq:deterministic} can be upper bounded exponentially with time~\cite{FB:22-CTDS}. 
If there exists $\alpha>0$ such that $c_t< -\alpha$ for all $t$, then the distance between two arbitrary trajectories of \eqref{eq:deterministic} is decreasing over time, and we say the system is \textit{contracting}~\cite{WL-JJES:98,FF-RS:14,ZA-EDS:14}. 
In practice, to apply the contraction theory for reachability analysis, one needs to estimate or bound $\mu(D_xf(x,u,t))$.
Several approaches have been proposed in the literature to determine the upper bound of $\mu(D_xf(x,u,t))$ (see e.g., \cite{chuchu2017simulation},\cite[Chapter 3,4]{FB:22-CTDS},\cite{EMA-PAP-JJES:08,ZZ:03}). These methods are applicable not only to contracting systems but also to systems with any $c_t\in\R$. In this paper, we allow $c_t\in\R$ rather than restricting it to be negative. 

\section{Reachability of Stochastic Systems} \label{sec: formulation}
In many real-world applications, the underlying dynamics are driven not only by deterministic inputs but also by stochastic disturbances. 
Existing methods and techniques for deterministic reachability analysis designed for deterministic and often bounded inputs/disturbances are not applicable to these scenarios with stochastic disturbances. 
We aim to bridge this gap by developing a unified framework of reachability analysis for stochastic systems. In this section, we formulate our probabilistic reachability problem and introduce our overall strategy for addressing it.

\subsection{Problem Statement}
Consider the stochastic system 
\begin{equation}\label{eq:stochastic}
    dX_t=f(X_t,u_t,t)dt+g_t(X_t)dW_t,
\end{equation}
where the state $X_t\in \R^n$ is a random vector, $u_t: \real_{\ge 0}\to\mU\subseteq \R^{p}$ is the input, $g_t$ is the diffusion coefficient, and $W_t\in\R^m$ is an $m$-dimensional Wiener process (Brownian motion) modeling the stochastic uncertainty. This stochastic system can be viewed as a noisy version of the deterministic system 
\begin{equation}\label{eq:associate-deterministic}
    \dot{x}_t=f(x_t,u_t,t).
\end{equation}
To ensure \eqref{eq:stochastic} has a solution, we default standard Lipschitz and linear growth conditions \cite[Theorem 5.2.1]{BO:13}. For reachability analysis, we impose the following assumption.
\begin{assumption}\label{as: boundness}
    For the stochastic system~\eqref{eq:stochastic}, there exist integrable curves $t\mapsto c_t$ and  $t\mapsto \sigma_t$ such that,
    \begin{enumerate}
        \item $\mu(D_xf(x,u,t))\leq c_t$ for any $t\ge 0$, $u\in\mU$, and $x\in\R^n$.
        \item $g_t(x)g_t(x)^{\top}\preceq \sigma_t^2 I_n$ for any $t\geq0$ and $x\in\R^n$.
    \end{enumerate}
\end{assumption}

\begin{figure}
\centering
\includegraphics[width =0.6\linewidth]{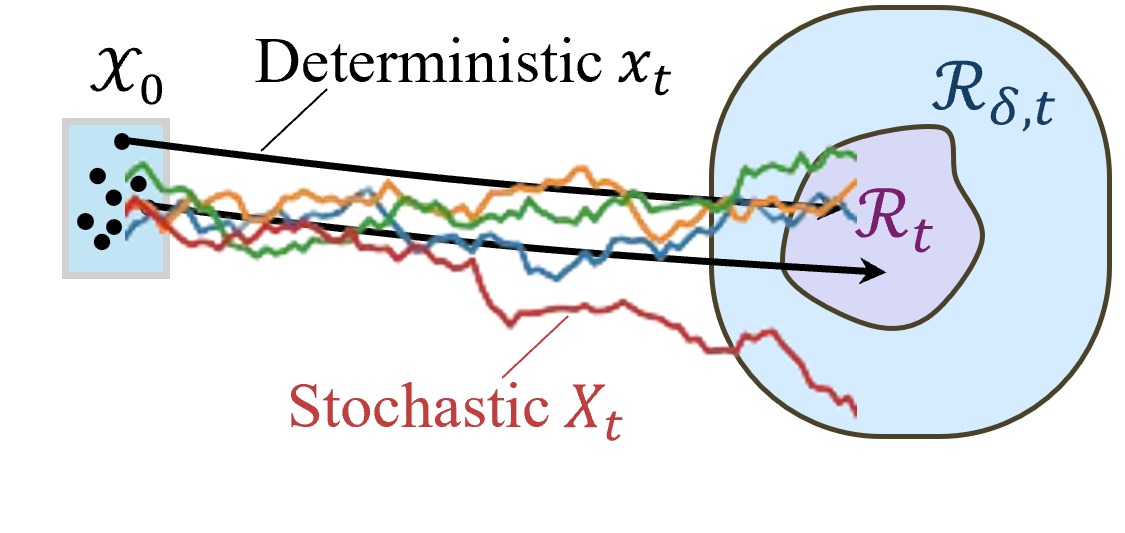}
\caption{An illustration of $\delta$-PRS at time $t$. Here $\mR_{\delta,t}$ is a $\delta$-PRS of the stochastic system \eqref{eq:stochastic}, whose trajectories are in color, and $\mR_t$ is the DRS of the associated deterministic system \eqref{eq:associate-deterministic}, whose trajectories are in black. 
}
\label{fig: prs}
\end{figure}

We are interested in characterizing the reachable set of the stochastic system \eqref{eq:stochastic} under Assumption \ref{as: boundness}. Departing from the deterministic dynamics \eqref{eq:associate-deterministic} driven only by the input $u_t$, the stochastic system \eqref{eq:stochastic} is driven by both the input $u_t$ and stochastic disturbance $dW_t/dt$. Deterministic reachability analysis falls short of capturing this stochastic disturbance. Indeed, most methods in deterministic reachability analysis assume bounded input/disturbance and approximate its DRS through worst-case type analysis \cite{cosner2024bounding}. However, the stochastic disturbance $dW_t/dt$ is unbounded \cite[Chapter 4.1]{sarkka2019applied}. This unbounded stochastic disturbance often results in a trivial reachable set in the sense of \eqref{eq:reach}. For example, the classical reachable set of the system $dX_t=dW_t$ is the entire state space for any $t>0$. 
We resort to a probabilistic notion of reachable sets to overcome these limitations of deterministic reachability analysis.
\begin{definition}[$\delta$-PRS]\label{def: p-PRS}
    Consider the stochastic system \eqref{eq:stochastic} with initial set $\mX_0\subseteq\R^n$ and input set $\mU\subseteq\R^p$. Given $\delta\in (0,1]$ and $t\ge 0$, the set $\mathcal{R}_{\delta,t}\subseteq \real^n$ is a $\delta$-probabilistic reachable set ($\delta$-PRS) of~\eqref{eq:stochastic} at time $t$, if for any $x_0\in\mX_0$ and piecewise continuous $u_t: \R_{\ge 0}\to\mU$, we have
    \begin{equation}\label{eq: def p-PRS}
        \prob{X_t\in\mR_{\delta,t}}\geq 1-\delta.
    \end{equation}
\end{definition}
Briefly, a probabilistic reachable set of a stochastic system~\eqref{eq:stochastic} is the set all possible trajectories can reach with high probability. 
An illustration of $\delta$-PRS is given in Figure \ref{fig: prs}. For sufficiently small $\delta$, $\mR_{\delta,t}$ contains the DRS of the associated deterministic system \eqref{eq:associate-deterministic} due to the stochastic disturbance, that is, $\mR_{t}\subseteq \mR_{\delta,t}$. 
By definition, the $\delta$-PRS is not unique. If $\mR_{\delta,t}$ is a $\delta$-PRS, then any $\mR_{\delta,t}'\supseteq \mR_{\delta,t}$ is also a $\delta$-PRS. We say $\mR_{\delta,t}$ is a \textit{tighter} $\delta$-PRS than $\mR_{\delta,t}'$ if $\mR_{\delta,t}\subseteq \mR_{\delta,t}'$.

In many applications involving reachability analysis, it is desirable to have a tight $\delta$-PRS. For instance, for safety-critical control, the safety of the system can be guaranteed by ensuring that the $\delta$-PRS does not overlap with the unsafe regions~\cite{cosner2023robust}. A loose $\delta$-PRS can result in very conservative strategies.
Therefore, we are interested in finding the tightest possible $\delta$-PRS.
\begin{problem}\label{problem: prs}
    Find an as tight as possible $\delta$-PRS $\mR_{\delta,t}$ of the stochastic system \eqref{eq:stochastic} under Assumption \ref{as: boundness}.
\end{problem}

\subsection{Separation Strategy and Stochastic Deviation}
The trajectories of the stochastic system \eqref{eq:stochastic} are driven by both deterministic input and stochastic disturbance/input. The effects of these two types of inputs on the trajectories are relatively independent and may be handled separately. Building on this intuition, we propose a strategy termed \textit{separation strategy} for probabilistic reachability analysis. The effects of the deterministic input can be encoded by the DRS of the associated deterministic system \eqref{eq:associate-deterministic}. To capture the effects of the stochastic disturbance, we associate each trajectory $X_t$ of the system \eqref{eq:stochastic} with a trajectory $x_t$ of the system \eqref{eq:associate-deterministic} with the same initial state $x_0=X_0$ and the same deterministic input $u_t$. The influence of the stochastic disturbance can then be represented by the deviation $\|X_t-x_t\|$. The probabilistic reachable set of \eqref{eq:stochastic} can be approximated by combining these two components as formalized below.
\begin{figure}
\centering
\includegraphics[width =0.9\linewidth]{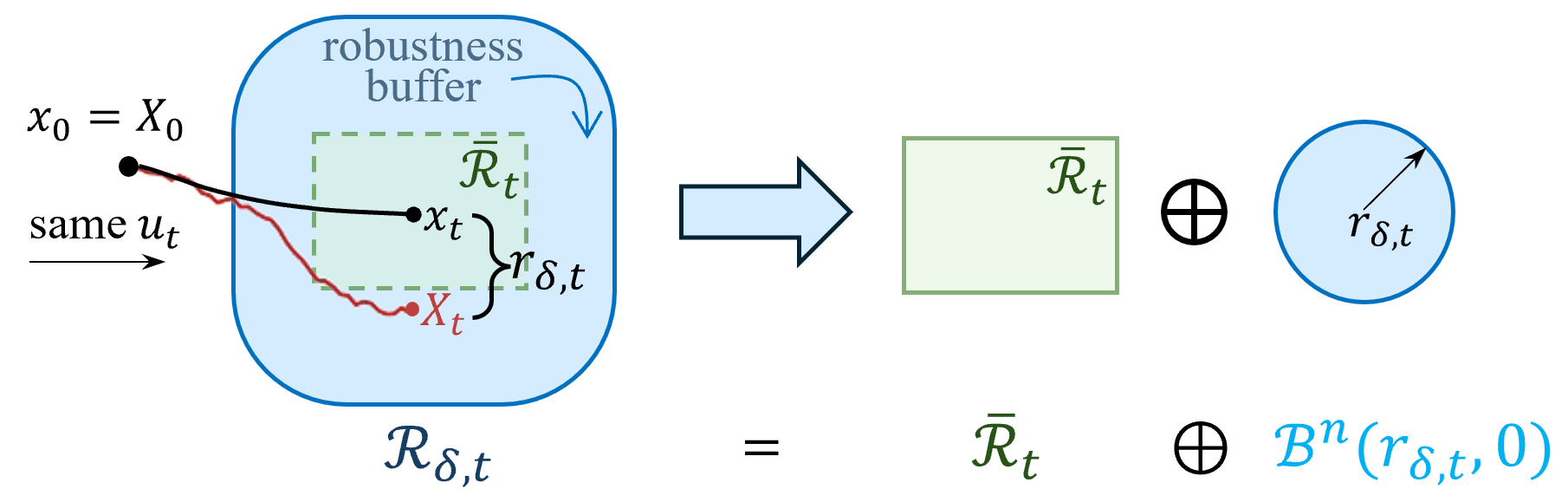}
\caption{An illustration of separation strategy. Here $\mR_{\delta,t}$ is a $\delta$-PRS of the stochastic system \eqref{eq:stochastic}, whose trajectory is $X_t$ in red. $\overline{\mR}_t$ is an over-approximation of the DRS of the associated deterministic system \eqref{eq:associate-deterministic}, whose trajectory is $x_t$ in black. The Minkowski sum corresponds to Proposition \ref{prop: separation}.}
\label{fig: separation}
\end{figure}

\begin{proposition}[Separation strategy]\label{prop: separation}
Consider the stochastic system \eqref{eq:stochastic} with its associated deterministic system \eqref{eq:associate-deterministic}.  
Let $\overline{\mathcal{R}}_t$ be any over-approximation of the DRS of \eqref{eq:associate-deterministic}. If there exists $r_{\delta,t}$ such that, for any given trajectory $x_t$ of \eqref{eq:associate-deterministic} and any associated trajectory $X_t$ of \eqref{eq:stochastic} with the same initial condition $x_0$ and input $u_\tau$, 
\begin{equation}\label{eq:rdt}
   \mathbb{P}\left(\|X_t-x_t\|\leq r_{\delta,t}\right)\geq 1-\delta,
   \end{equation} 
then $\overline{\mR}_t\oplus\mathcal{B}^n(r_{\delta,t},0)$ is a $\delta$-PRS of \eqref{eq:stochastic}.
\end{proposition}
\begin{proof}
Let $X_t$ be any trajectory of \eqref{eq:stochastic} associated with a trajectory $x_t$ of \eqref{eq:associate-deterministic}, then, by the assumption \eqref{eq:rdt} and the definition of the Minkowski sum~\cite{fogel2007exact},
\begin{align}\nonumber
    X_t\in \{x_t\}\oplus\mathcal{B}^n(r_{\delta,t},0)
\end{align}
with probability at least $1-\delta$. By the definition of $\overline{\mathcal{R}}_t$, $x_t\in\overline{\mathcal{R}}_t$. Therefore, with probability at least $1-\delta$,
    \[
        X_t\in\overline{\mR}_t\oplus\mathcal{B}^n(r_{\delta,t},0),
    \]
which completes the proof.
\end{proof}

We term the difference $\|X_t-x_t\|$ between associated trajectories {\em stochastic deviation}. A key ingredient of Proposition \ref{prop: separation} is a probabilistic bound $r_{\delta,t}$ that upper bounds the stochastic deviation with high probability. Proposition \ref{prop: separation} states that if a probabilistic bound $r_{\delta,t}$ exists, then the dilation of the reachable set of the deterministic system \eqref{eq:associate-deterministic} with a ball of radius $r_{\delta,t}$ is a $\delta$-PRS of \eqref{eq:stochastic}. This separation strategy decomposes the probabilistic reachability analysis problem into two parts: approximate the DRS of \eqref{eq:associate-deterministic} and estimate the probabilistic bound $r_{\delta,t}$ of the stochastic deviation.
Once a bound $r_{\delta,t}$ of the stochastic deviation is provided, one can combine it with any existing deterministic reachability method to approximate the $\delta$-PRS. 

The size of the $\delta$-PRS $\overline{\mR}_t\oplus\mathcal{B}^n(r_{\delta,t},0)$ in Proposition \ref{prop: separation} increases with $r_{\delta,t}$. To ensure $\overline{\mR}_t\oplus\mathcal{B}^n(r_{\delta,t},0)$ is not an overly-conservative $\delta$-PRS of \eqref{eq:stochastic}, it is crucial to establish an as tight as possible probabilistic bound $r_{\delta,t}$ for the stochastic deviation. This is the main challenge addressed in this paper.
\begin{problem}\label{problem: vib}
Establish an as tight as possible probabilistic bound $r_{\delta,t}$ of the stochastic deviation $\|X_t-x_t\|$ associated with systems \eqref{eq:stochastic}-\eqref{eq:associate-deterministic} under Assumption \ref{as: boundness}.
\end{problem}

\section{Expectation Bound and Limitations}\label{sec: limit}
To warm up, we first revisit an existing approach~\cite{QCP-NT-JJS:09} for Problem \ref{problem: vib} and highlight its limitations. 
\subsection{Expectation Bound on Stochastic Deviation}
Inspired by \cite{QCP-NT-JJS:09} we present a method that probabilistically bounds the stochastic deviation $\|X_t-x_t\|$ under Assumption \ref{as: boundness} by bounding the expectation $\mbE(\|X_t-x_t\|^2)$.  

For a trajectory $X_t$ of the stochastic system \eqref{eq:stochastic} and the associated trajectory $x_t$ of the deterministic system \eqref{eq:associate-deterministic}, define the Lyapunov function $V_t = \|X_t-x_t\|^2$. Then a direct application of the Ito's Lemma \cite{sarkka2019applied} yields
    \begin{align}\nonumber
    dV_t & = 2\left(X_t-x_t\right)^{\top}(f(X_t,u_t,t)-f(x_t,u_t,t)) dt \\ & + \mathrm{tr}(g_t^{\top}g_t) dt + 2 (X_t-x_t)^{\top}g_tdW_t \label{eq: dV_t Ito}
    \end{align}
Following standard It\'{o}  Calculus, for every $t,h\ge 0$,
\begin{align*}
   \mathbb{E}(V_{t+h})&- \mathbb{E}(V_t) =  \mathbb{E}\left(\int_{t}^{t+h} dV_s \right) \\& \le \int_{t}^{t+h} \mathbb{E}(dV_s)\\&  \le  \int_{t}^{t+h} (2c_s \mathbb{E}(\|X_s-x_s\|^2) + n\sigma^2_s) ds  \\&
   =  \int_{t}^{t+h} \left(2c_s \mathbb{E} (V_s) + n\sigma^2_s\right) ds.
\end{align*}
where the first inequality holds by the triangle inequality and the second inequality holds by  Lemma~\ref{lemma: matrix measure}. Taking the limsup of both side as $h\to 0$, for every $t\ge 0$, we get
\begin{align}\label{eq: dE}
    D^{+}\mathbb{E}(V_t) \le 2c_t\mathbb{E}(V_t) + n\sigma_t^2, \quad V_0=0,
\end{align}
where $D^+$ is the upper Dini Derivative with respect to $t$. By the generalized Gr\"{o}wall-Bellman lemma~\cite[Appendix A1, Proposition 4]{TL:10}, it follows the expectation bound
\begin{equation} \label{eq: E-bound}
    \mbE(\|X_t-x_t\|^2)=\mbE(V_t)\leq n\Psi_t, 
\end{equation}
where 
    \begin{subequations}\label{eq:Psi}
    \begin{eqnarray}
        \Psi_t&=&e^{2\psi_t}\int_0^t \sigma_\tau^2e^{-2\psi_\tau}d\tau
    \\
    \psi_t&=&\int_0^t c_\tau d\tau.
    \end{eqnarray}
    \end{subequations}
Applying Markov inequality to the expectation bound \eqref{eq: E-bound}, we obtain the probabilistic bound
\begin{equation}\label{eq:HBP}
        \mathbb{P}\left(\|X_t - x_t\| \le  \sqrt{\frac{n}{\delta}\Psi_t}\right) 
        = \prob{V_t \leq \frac{n}{\delta}\Psi_t}\geq 1-\delta
\end{equation}
for any $\delta\in(0,1)$.

\subsection{Limitations of Expectation Bound}\label{subsec: limit}
The bound \eqref{eq:HBP} based on the expectation bound \eqref{eq: E-bound} turns out to be loose. 
To see this, consider the linear time-invariant (LTI) stochastic system
\begin{align}\label{eq:LS}
    d X_t = (A X_t + B u_t)dt + \sigma dW_t
\end{align}
and the associated deterministic system
\begin{align}\label{eq:LD}
    \dot{x}_t = A x_t + B u_t. 
\end{align}
In this case, the bound \eqref{eq:HBP} reads
\begin{equation}\label{eq:bound1}
    \mathbb{P}\left(\|X_t - x_t\| \le r_{\delta,t}^{(1)} \right) \ge 1-\delta,
\end{equation}
where $r_{\delta,t}^{(1)}=\sqrt{\frac{n\sigma^2(e^{2ct}-1)}{2c\delta}}$ with $c=\mu(A)$. 

On the other hand, when initialized at $X_0=x_0$, $X_t$ is a Gaussian random variable \cite{sarkka2019applied} with mean $\mbE(X_t) = x_t$ and covariance matrix
\begin{align}\label{eq: cov(X_t)}
    \cov{X_t}=\int_{0}^{t} \sigma^2e^{A(t-\tau)} e^{A^{\top}(t-\tau)}d\tau.
\end{align}
Invoking the fact that $\|e^{At}\|\leq e^{\mu(A)t}$ for any $t\geq0$ \cite{FB-MF-EH:74}, $\cov{X_t}$ can be bounded as
\begin{equation}\label{eq: bound cov(X_t)}
    \begin{split}
        \cov{X_t} &\preceq\int_{0}^{t} \sigma^2\|e^{A(t-\tau)}\| \|e^{A^{\top}(t-\tau)}\| d\tau \, I_n \\ & \preceq
    \sigma^2 \int_{0}^{t} e^{2c(t-\tau)}d\tau \, I_n\\ & = 
    \tfrac{\sigma^2}{2c}(e^{2ct}-1)\, I_n.
    \end{split}
\end{equation}
By the concentration property of Gaussian distribution \cite[Chapter 7]{gittens2011tail}, for any $\delta\in(0,1)$, with probability at least $1-\delta$, 
\begin{equation}\label{eq: Gaussian tail}
    \|X_t-x_t\|\le \sqrt{\|\cov{X_t}\|}(4\sqrt{n} + 2\sqrt{2\log(1/\delta)}).
\end{equation}
Plugging \eqref{eq: bound cov(X_t)} into \eqref{eq: Gaussian tail} yields
\begin{align}\label{eq:bound2}
    \mathbb{P}\left(\|X_t-x_t\|\le r^{(2)}_{\delta,t}\right) \ge 1-\delta,
\end{align}
where $r_{\delta,t}^{(2)}=\sqrt{\tfrac{\sigma^2}{2c}(e^{2ct}-1)}(4\sqrt{n} + 2\sqrt{2\log(1/\delta)})$. 
\smallskip

The bound \eqref{eq:bound2} is substantially better than \eqref{eq:bound1}.
While the dependency of $r_{\delta,t}^{(1)}$ and $r_{\delta,t}^{(2)}$ on $c$ and $n$ are of the same order, the dependency of $r^{(2)}_{\delta,t}$ on $\delta$ is $\mO(\sqrt{\log (1/\delta)})$, much better than the $\mathcal{O}\left(\sqrt{1/\delta}\right)$ dependency of $r^{(1)}_{\delta,t}$ on $\delta$. For small $\delta$ (e.g., $10^{-10}$), which is crucial for safety-critical systems, $\sqrt{\log (1/\delta)}$ is significantly smaller than $\sqrt{1/\delta}$ ($4.80$ v.s. $10^5$). As a result, the probabilistic reachable set based on the bound \eqref{eq:HBP} can be conservative in practice. 

Thus, there is a significant gap between the result \eqref{eq:HBP} for nonlinear dynamics and probabilistic bounds for linear dynamics. 
The limitation of the expectation bound primarily lies in the quadratic Lyapunov function $V_t=\|X_t-x_t\|^2$. The analysis focuses only on the evolution of \textit{the second order moment} $\mathbb{E}(\|X_t-x_t\|^2)$. It can at best guarantee a probabilistic bound for $\|X_t-x_t\|$ of order $\mO(\sqrt{1/\delta})$ via Markov inequality. This gives rise to the question: is the gap fundamental or an artifact of the analysis?

\section{Probabilistic Bound on Stochastic Deviation}\label{sec:propag}
In this section, we answer the aforementioned question by establishing a probabilistic bound for the stochastic deviation $\|X_t-x_t\|$ of order $\mO(\sqrt{\log (1/\delta)})$ for general nonlinear stochastic systems \eqref{eq:stochastic} under Assumption \ref{as: boundness}. We further show our bound is consistent with that for linear systems under the same assumption and is thus tight.

\subsection{Sub-Gaussian and MGF}\label{subsec: motivation}
The analysis \eqref{eq: cov(X_t)}-\eqref{eq:bound2} relying on the Gaussianity for linear systems can not be applied to \eqref{eq:stochastic} since $X_t$ is not necessarily Gaussian for nonlinear systems.
Fortunately, the norm concentration property \eqref{eq: Gaussian tail} holds not only for Gaussian random vectors (distributions) but also for a wider class of random vectors known as sub-Gaussian vectors (distributions). 
\begin{definition} \label{def: subG}
    A random variable $X\in\R^n$ is said to be sub-Gaussian with variance proxy $\sigma^2$, denoted as $X\sim subG(\sigma^2)$, if $\mbE_X(X)=0$ and
\begin{equation}\label{eq: subG}
    \mbE_X\left(e^{\lambda \innerp{\ell,X}}\right)\leq e^{\frac{\lambda^2\sigma^2}{2}},~\forall \lambda\in\R, ~\forall \ell\in\mS^{n-1}.
\end{equation}
\end{definition}
Many distributions including Gaussian distribution, zero-mean uniform distribution, and any zero-mean distribution with bounded support are instances of sub-Gaussian distributions.
For Gaussian distribution, the variance proxy $\sigma^2$ is $\|\cov{X}\|$. 

Sub-Gaussian distributions share the same norm concentration property as Gaussian distributions. For the sake of completeness, we present a version of the concentration property and its proof in Appendix \ref{app: pf norm conc}.
\begin{lemma}\label{lemma: concentration}
    Let $X\in\R^n$ be a sub-Gaussian random vector with variance proxy $\sigma^2$, then for any $\delta\in(0,1)$ and any $\varepsilon\in(0,1)$, 
    \begin{equation}\label{eq: concentration norm}
        \|X\|\leq \sigma\sqrt{\varepsilon_1n +\varepsilon_2\log(1/\delta)}
    \end{equation}
holds with probability at least $1-\delta$, where
    \begin{equation}\label{eq: epsilon val}
        \varepsilon_1=\frac{2\log(1+2/\varepsilon)}{(1-\varepsilon)^2},~ \varepsilon_2=\frac{2}{(1-\varepsilon)^2}.
    \end{equation}
\end{lemma}

Lemma \ref{lemma: concentration} states a probabilistic bound of the norm $\|X\|$ of a sub-Gaussian random vector that scales as $\mathcal{O}(\sqrt{n})$ and $\mathcal{O}(\sqrt{\log(1/\delta)})$, the same as \eqref{eq: Gaussian tail}. The parameter $\varepsilon$ can be selected according to the values of $n, \delta$ to minimize the bound. When $\varepsilon=0.5$, $\varepsilon_1=8\log5 \approx 16$ and $\varepsilon_2=8$. Since $\sigma^2=\|\cov{X}\|$ for Gaussian, \eqref{eq: concentration norm} becomes \eqref{eq: Gaussian tail} after applying Jensen's Inequality. The dependence $\mathcal{O}(\sqrt{n})$ and $\mathcal{O}(\sqrt{\log(1/\delta)})$ in Lemma \ref{lemma: concentration} is tight, but the expressions of $\varepsilon_1, \varepsilon_2$ in \eqref{eq: epsilon val} are constructed in the proof and are by no means optimal, especially for specific values of $n$. For example, when the dimension $n=1$, by Hoeffding's Inequality \cite[Chapter 1.2]{rigollet2023high}, a better choice is $\varepsilon_1=2\log2$ and $\varepsilon_2=2$.

To show a random variable $X$ is sub-Gaussian, one needs to verify $\mbE_X(X)=0$ and the inequality \eqref{eq: subG}. Note that the left-hand side of \eqref{eq: subG} is the Moment Generating Function (MGF) \cite[Chapter 1.1]{rigollet2023high}
\begin{equation}\label{eq: MGF}
    \mbE_X\left(M_{\lambda,\ell}(X)\right)\defeq \mbE_X\left(e^{\lambda\innerp{\ell,X}}\right),\quad \ell\in\mS^{n-1}
\end{equation}
a common tool for concentration analysis. One advantage of the MGF compared with the second-order moment used in Section \ref{sec: limit} is that the MGF captures high-order information, and this is a major reason why MGF is useful for analyzing concentration properties. 

Thus, a potential approach to bound $\|X_t-x_t\|$ is to show $X_t-x_t$ is sub-Gaussian. Unfortunately, this is not true. For associated trajectories $X_t$ and $x_t$, $\mbE(X_t)\neq x_t$ for general nonlinear dynamics \cite[Chapter 5.5]{sarkka2019applied}. Moreover, \eqref{eq: subG} requires bounding the evolution of the MGF for all $\ell\in\mS^{n-1}$, which can be too strong.

\subsection{Averaged Moment Generating Function}\label{sub:AMGF}
Inspired by the concentration properties of sub-Gaussian distributions and the limitations of MGF, we propose a weaker version of MGF termed the \textit{Averaged Moment Generating Function (AMGF)} for probabilistic reachability analysis. 

\begin{definition}[AMGF]\label{def: salf}
Given $\lambda\in\R$, the Averaged Moment Generating Function $\Phi_{n,\lambda}: \R^n\to\R$ is defined as
    \begin{equation}\label{eq: AMGF}
        \mbE_X(\Phi_{n,\lambda}(X))\defeq \mbE_X\expectw{\ell\sim\mS^{n-1}}{e^{\lambda\innerp{\ell,X}}}.
    \end{equation}
\end{definition}

The AMGF is an average of the MGF over the sphere $\ell\sim\mS^{n-1}$. It was recently proposed in \cite{altschuler2022concentration} to study sampling problems. Thanks to the averaging, bounding the AMGF is easier than bounding MGF for each $\ell$. The AMGF can also be viewed as an MGF by replacing the exponential energy function $e^{\lambda\innerp{\ell,x}}$ by $\Phi_{n,\lambda}(x)=\expectw{\ell\sim\mS^{n-1}}{e^{\lambda\innerp{\ell,x}}}$. This energy function $\Phi_{n,\lambda}$ has several intriguing properties. 
\begin{lemma}[Properties of $\Phi_{n,\lambda}$]\label{lemma: AMGF_1}
    The following statements hold for $\Phi_{n,\lambda}$ in \eqref{eq: AMGF}: 
    \begin{enumerate}
        \item\label{p1:AMGF} {\it Rotation invariance:} For any $x\in\R^n$ and $\eta\in\mS^{n-1}$, 
        \begin{align*}
            \salf{x} = \salf{\|x\| \eta}.
        \end{align*}
        \item\label{p2:AMGF} {\it Monotonicity:} For any $x,y\in\R^n$ such that $\|x\|\leq\|y\|$, 
        $$ 1\le\salf{x}\leq \salf{y}.$$
    \end{enumerate}
\end{lemma}
\smallskip

Lemma \ref{lemma: AMGF_1} implies that $\salf{x}$ only depends on the norm $\|x\|$ of $x$ and is monotonically increasing as $\|x\|$. For a non-expanding deterministic system \eqref{eq:associate-deterministic}, that is, $\mu(D_xf(x,u,t)) \le 0$, these properties imply that $\salf{x_t-y_t}$ is decreasing for any two arbitrary trajectories $x_t, y_t$. This can be formalized as follows.
\begin{lemma}\label{lemma: exp.<>}
      Consider the deterministic system \eqref{eq:associate-deterministic} such that $\mu(D_xf(x,u,t)) \le 0$ for every $(x,u,t)\in\real^n\times\mU\times\real_{\ge 0}$, then for any $x,y\in\R^n$, $u\in\mU$ and $t\geq0$:
        \begin{align*}
            \mathbb{E}_{\ell\sim \mS^{n-1}} \left(e^{\lambda\langle \ell, x-y\rangle}\lambda\ell^{\top}(f(x,u,t)-f(y,u,t))\right) \le 0. 
        \end{align*}
\end{lemma}

An intriguing fact about AMGF is that it induces the same concentration property as MGF.
\begin{lemma}\label{lemma: salf to sg}
    If a random variable $X\in\R^n$ satisfies
    \begin{equation}\label{eq: lemma salf_2}
    \mbE_X\left(\salf{X}\right)\leq e^{\frac{\lambda^2\sigma^2}{2}},~\forall \lambda\in\R,
\end{equation}
    then for any $\delta>0$, \eqref{eq: concentration norm} holds with probability at least $1-\delta$.
\end{lemma}
At first sight, this is counter-intuitive, since upper-bounding AMGF is weaker than upper-bounding MGF for all $\ell$. To see why Lemma \ref{lemma: salf to sg} holds, define an intermediate random variable $\tX=QX$ where $Q\sim\mathbb{U}^n$ is a random unitary matrix with $\mathbb{U}^n$ denoting the set of all the unitary matrices in $\R^{n\times n}$. Then the AMGF over $X$ is equal to the MGF over $\tX$, that is, 
$\mbE_X\left(\salf{X}\right)=\mbE_{\tX}\left(e^{\lambda\innerp{\ell,\tX}}\right)$. 
This means $\tX$ is sub-Gaussian with variance proxy $\sigma^2$. Lemma \ref{lemma: salf to sg} then follows by noticing that the transformation $\tX=QX$ does not affect the norm.

\subsection{Theoretical Analysis}\label{sec:theory}
Equipped with the AMGF, we are ready to establish a tighter probabilistic bound for the stochastic deviation $\|X_t-x_t\|$. Thanks to Lemma \ref{lemma: salf to sg}, it suffices to bound the evolution of the AMGF $\mbE(\salf{X_t-x_t})$ over time. Below we establish a probabilistic bound of the stochastic deviation of order $\mO(\sqrt{\log (1/\delta)})$ for the stochastic system \eqref{eq:stochastic} satisfying Assumption \ref{as: boundness} by developing a tight bound of $\mbE(\salf{X_t-x_t})$.
\begin{theorem}\label{thm: sto vib}
    Consider the stochastic system \eqref{eq:stochastic} and the deterministic system \eqref{eq:associate-deterministic} under Assumption \ref{as: boundness}. Let $X_t$ be a trajectory of \eqref{eq:stochastic} and $x_t$ be an associated trajectory of \eqref{eq:associate-deterministic} with the same initial condition $x_0$ and input $u_t:t\to\mU$. Then, for any $t>0$, $\delta\in(0,1)$ and $\varepsilon\in(0,1)$,
    \begin{equation} \label{eq: thm1}
         \|X_t-x_t\|\leq \sqrt{\Psi_t(\varepsilon_1n+\varepsilon_2\log(1/\delta))},
     \end{equation}
     holds with probability at least $1-\delta$,
     where $\Psi_t$ is as in \eqref{eq:Psi} and $\varepsilon_1$,$\varepsilon_2$ are given by \eqref{eq: epsilon val}.
\end{theorem}

\begin{proof}
We start with a special case where Assumption \ref{as: boundness} holds with a global matrix measure bound $c_t=0$ and then generalize it to cases where Assumption \ref{as: boundness} holds with arbitrary $c_t$.

\subsubsection{Special Case}
Denote $v_t=X_t-x_t$ and $\beta_t=f(X_t,u_t,t)-f(x_t,u_t,t)$, then
\begin{equation}\label{eq: Xt-xt sde}
     dv_{t}= \beta_tdt+ g_tdW_t.
\end{equation} 
Based on the Fokker–Planck equation \cite{sarkka2019applied}, $h_t=\mbE(\salf{v_t})$ satisfies 
\begin{equation}\label{eq: FPK Eh(t)}
    \frac{dh_t}{dt}=\expect{\innerp{\nabla\salf{v_t},\beta_t}} 
        +\tfrac{1}{2}\expect{\innerp{\nabla^2\salf{v_t},g_tg_t^{\top}}}
\end{equation}
By \eqref{eq: AMGF}, 
\begin{equation}\label{eq: cal part 1 FPK}
        \expect{\innerp{\nabla\salf{v_t},\beta_t}} 
        = \mbE\,\mbE_{\ell\sim\mS^{n-1}} \left(e^{\lambda\innerp{\ell,v_t}}\lambda\ell^{\top}\beta_t \right).
\end{equation}
Applying Lemma \ref{lemma: exp.<>} with $x=X_t$ and $y=x_t$, we obtain
\begin{equation}\label{eq: apply lemma exp<>}
    \mbE_{\ell\sim\mS^{n-1}}\left(e^{\lambda\innerp{\ell,v_t}}\lambda\ell^{\top}\beta_t\right)\leq 0.
\end{equation}
Then 
\begin{equation}\label{eq: part 1 FPK}
    \expect{\innerp{\nabla\salf{v_t},\beta_t}}\leq 0
\end{equation}
follows by taking the expectation of \eqref{eq: apply lemma exp<>}.

The term $\frac{1}{2}\expect{\innerp{\nabla^2\salf{v_t},g_tg_t^{\top}}}$ can be bounded as 
\begin{equation}\label{eq: part 2 FPK}
    \begin{split}
        \tfrac{1}{2}&\expect{\innerp{\nabla^2\salf{v_t},g_tg_t^{\top}}} \\
        =&\tfrac{1}{2}\mbE\, \expectw{\ell\sim\mS^{n-1}}{\innerp{\lambda^2e^{\lambda\innerp{\ell,v_t}}\ell\ell^{\top},g_tg_t^{\top}}} \\
        \leq& \tfrac{1}{2}\mbE\,\expectw{\ell\sim\mS^{n-1}}{\lambda^2e^{\lambda\innerp{\ell,v_t}}\tr(\ell\ell^{\top})\,\|g_tg_t^{\top}\|} \\
        \leq& \frac{\lambda^2\sigma_t^2}{2}\expect{\salf{v_t}}=\frac{\lambda^2\sigma_t^2}{2}h_t
    \end{split}
\end{equation}
where the first inequality follows the Cauchy–Schwarz inequality and the last line uses the fact that $\tr(\ell\ell^{\top})=1$ for any $\ell\sim\mS^{n-1}$ and $\|g_tg_t^{\top}\|\leq \sigma_t^2$ as in Assumption \ref{as: boundness}.

Plugging \eqref{eq: part 1 FPK} and \eqref{eq: part 2 FPK} into \eqref{eq: FPK Eh(t)} we arrive at
\begin{equation}\label{eq: ode c=0}
    \frac{dh_t}{dt}\leq \frac{\lambda^2\sigma_t^2}{2}h_t,\quad h_0=1,
\end{equation}
and using the Gr\"{o}nwall inequality \cite{gronwallEq}, we conclude 
\begin{equation}\label{eq: E(Phi)<=}
        \expect{\Phi_{n,\lambda}(X_t-x_t)} = h_t\leq e^{\frac{\lambda^2\int_0^t\sigma_\tau^2d\tau}{2}}.
    \end{equation}    
By Lemma \ref{lemma: salf to sg}, \eqref{eq: E(Phi)<=} implies that, for $\forall \delta\in(0,1)$, with probability at least $1-\delta$, 
    \begin{equation}\label{eq: result c=0}
         \|X_t-x_t\|\leq \sqrt{(\varepsilon_1n+\varepsilon_2\log(1/\delta))\int_0^t\sigma_\tau^2d\tau},
     \end{equation}
where $\varepsilon_1,\varepsilon_2$ satisfy \eqref{eq: epsilon val}. Since $\Psi_t=\int_0^t \sigma_\tau^2 d\tau$ when $c_t=0$ by definition, \eqref{eq: result c=0} corresponds to \eqref{eq: thm1}. 
This completes the proof in the special case.

\subsubsection{General Cases}
Next we consider the general cases where Assumption \ref{as: boundness} holds with $\mu(D_xf(x,u,t))\leq c_t$ for arbitrary $c_t\in\R$. The strategy is to convert them into the above special case via scaling. 
Define scaled trajectories $\tX_t=e^{-\psi_t}X_t$ and $\tx_t=e^{-\psi_t}x_t$ where $\psi_t=\int_0^t c_\tau d\tau$, then $\tx_t$ is a trajectory of the deterministic system
\begin{equation}\label{eq:associate-deterministic_tx}
    \dot{\tx}_t=-c_t\tx_t+e^{-\psi_t}f(e^{\psi_t}\tx_t,u_t,t)=: \tilde{f}(\tx_t,u_t,t),
\end{equation}
Similarly, $\tX_t$ satisfies 
\begin{equation}\label{eq:stochastic_tx}
        d\tX_t
        =\tilde{f}(\tX_t,u_t,t)dt+e^{-\psi_t}g_tdW_t.
\end{equation}

Note that \eqref{eq:stochastic_tx} and \eqref{eq:associate-deterministic_tx} have the same drift dynamics $\tilde f$. For any $\tx_t,\ty_t\in\R^n$, $\tilde f$ satisfies
\begin{equation*}
    \begin{split}
        &(\tx_t-\ty_t)^{\top}\left(\tilde{f}(\tx_t,u_t,t)-\tilde{f}(\ty_t,u_t,t)\right) \\
         =&(\tx_t-\ty_t)^{\top}\left(-c_t(\tx_t-\ty_t)+e^{-\psi_t}\left(f(x_t,u_t,t)-f(y_t,u_t,t)\right)\right) \\
        =& -c_t\|\tx_t-\ty_t\|^2+e^{-2\psi_t}(x_t-y_t)^{\top}\left(f(x_t,u_t,t)-f(y_t,u_t,t)\right) \\
        \leq & -c_t\|\tx_t-\ty_t\|^2+e^{-2\psi_t} c_t\|x_t-y_t\|^2 \\
        =& -c_t\|\tx_t-\ty_t\|^2+c_te^{-2\psi_t} e^{2\psi_t}\|\tx_t-\ty_t\|^2 =0,  
    \end{split}
\end{equation*}
meaning Assumption \ref{as: boundness} holds for scaled systems \eqref{eq:associate-deterministic_tx} and \eqref{eq:stochastic_tx} with $\tilde{c}_t=0$ and the results for the special case can be applied. The diffusion coefficient of \eqref{eq:stochastic_tx} satisfies $\|e^{-2\psi_t}g_tg_t^{\top}\|\leq e^{-2\psi_t}\sigma_t^2=: \tilde{\sigma}_t^2$. Applying \eqref{eq: E(Phi)<=} to the scaled dynamics \eqref{eq:stochastic_tx} we have that with probability at least $1-\delta$,
    \begin{equation}\label{eq: result tilde_c=0}
         \|\tX_t-\tx_t\|\leq \sqrt{(\varepsilon_1n+\varepsilon_2\log(1/\delta))\int_0^t \tilde\sigma_\tau^2d\tau}.
     \end{equation}
Recalling $X_t=e^{\psi_t}\tX_t$, $x_t=e^{\psi_t}\tx_t$, and $\Psi_t$ in \eqref{eq:Psi}, we conclude that with probability at least $1-\delta$,
 \begin{equation*}\label{eq: result c>0}
 \begin{split}
      \|X_t-x_t\|\leq &\sqrt{(\varepsilon_1n+\varepsilon_2\log(1/\delta))e^{2\psi_t}\int_0^t \sigma_\tau^2e^{-2\psi_\tau}d\tau} \\
         =&\sqrt{\Psi_t(\varepsilon_1n+\varepsilon_2\log(1/\delta))},
 \end{split}
     \end{equation*}
which completes the proof.
\end{proof}

\begin{remark}
    When Assumption \ref{as: boundness} holds with time-invariant $c_t\equiv c$ and $\sigma_t\equiv\sigma$, $\psi_t$ defined in \eqref{eq:Psi} becomes $\psi_t=ct$, and \eqref{eq: thm1} in Theorem \ref{thm: sto vib} reduces to
    \begin{equation}\label{eq:timeinvariant}
    \|X_t-x_t\|\leq \sqrt{\frac{\sigma^2(e^{2ct}-1)}{2c}(\varepsilon_1n+\varepsilon_2\log(1/\delta))}.
\end{equation}
\end{remark}
The probabilistic bound in Theorem \ref{thm: sto vib} highly relies on the contraction rate of the dynamics. The bound \eqref{eq: thm1} and \eqref{eq:timeinvariant} resemble the input-to-state bounds used in contraction-based reachability of deterministic systems \cite{AD-SJ-FB:20o}. Thus, our results can be viewed as the stochastic counterpart of the deterministic incremental input-to-state bounds in contraction theory.

\subsection{Extension to Weighted Norm}\label{sec:weighted}

The probabilistic bound of the stochastic deviation $\|X_t-x_t\|$ in Theorem \ref{thm: sto vib} can be extended to bound the weighted deviation $\|X_t-x_t\|_P$ for any positive-definite matrix $P$.
To this end, define the weighted matrix measure of a matrix $A$ as
\begin{equation*}
        \mu_P(A)=\lim_{\epsilon\to 0^{+}}\frac{\|I_n + \epsilon A\|_P-1}{\epsilon},
    \end{equation*}
    which can be obtained using the expression $\mu_P(A) = \mu(P^{\frac{1}{2}}AP^{-\frac{1}{2}})$~\cite{FB:22-CTDS}. Consider the systems \eqref{eq:stochastic} and \eqref{eq:associate-deterministic} satisfying a modified version of Assumption \ref{as: boundness} as $\mu_P(D_xf(x,u,t))\leq c_t$ and $P^{\frac{1}{2}}g_tg_t^{\top}P^{\frac{1}{2}}\preceq \sigma_t^2 I_n$.

This setting with weighted norm can be converted to the unweighted version in Section \ref{sec:theory} through a coordinate transformation. More specifically, given associated trajectories $X_t, x_t$ of \eqref{eq:stochastic} and \eqref{eq:associate-deterministic}, define $\hat{X}_t=P^{\frac{1}{2}}X_t$, $\hat{x}_t=P^{\frac{1}{2}}x_t$, then $\hat{X_t}$ and $\hat{x}_t$ satisfy 
    \begin{align}
        d\hat{X}_t&=\hat{f}(\hat{X}_t,u_t,t)dt+\hat{g}_tdW_t, \label{sys: Pss}\\
        \dot{\hat{x}}_t&=\hat{f}(\hat{x}_t,u,t)\label{sys: Pds}
    \end{align}
with $\hat{f}(\hat{x})=P^{\frac{1}{2}}f(P^{-\frac{1}{2}}\hat{x})$ and $\hat{g}_t=P^{\frac{1}{2}}g_t$.    
By definition, $\mu_P(A)=\mu(P^{\frac{1}{2}}AP^{-\frac{1}{2}})$ for any matrix $A$, thus $\mu(D_{\hat{x}}\hat f(\hat{x},u,t))=\mu_P(D_xf(x,u,t))\leq c_t$. Moreover, $\hat{g}_t\hat{g}_t^{\top}= P^{\frac{1}{2}}g_tg_t^{\top}P^{\frac{1}{2}} \preceq \sigma_t^2 I_n$. Therefore, the systems \eqref{sys: Pss} and \eqref{sys: Pds} satisfy Assumption \ref{as: boundness} with the standrad $\ell_2$-norm. Then, by Theorem \ref{thm: sto vib}, with probability at least $1-\delta$,
    \begin{equation}\label{eq: P weighted sto vib}
            \|X_t-x_t\|_P=\|\hat{X}_t-\hat{x}_t\| 
            \leq \sqrt{\Psi_t(\varepsilon_1n+\varepsilon_2\log(1/\delta))}.
    \end{equation}

This extension for weighted norm can sometimes be advantageous to establish a tighter bound. 
Given a matrix $A$, $\mu(A)$ can be much larger than the real parts of the eigenvalues of $A$. In contrast, with a proper positive-definite matrix $P$, $\mu(P^{\frac{1}{2}}AP^{-\frac{1}{2}})$ can be made arbitrarily close to the real parts of the eigenvalues of $A$ \cite[Chapter 2.7]{FB:22-CTDS}. In this circumstance, working with the weighted norm can lead to sharper results.

\subsection{Tightness of Probabilistic Bound}\label{subsec: tightness}
Finally, we show that the probabilistic bound in Theorem \ref{thm: sto vib} is tight under Assumption \ref{as: boundness} and it is impossible to achieve better probabilistic bounds than \eqref{eq: thm1} without additional assumptions. In particular, we show that the bound \eqref{eq: thm1} precisely captures the stochastic deviation of linear systems satisfying Assumption \ref{as: boundness}.

Consider the LTI stochastic system \eqref{eq:LS} and its associated deterministic system \eqref{eq:LD}. They satisfy Assumption \ref{as: boundness} with $c_t\equiv c=\mu(A)$ and $\sigma_t \equiv \sigma$. By Theorem \ref{thm: sto vib}, with probability at least $1-\delta$,
\begin{equation}\label{eq: exam tight}
    \|X_t-x_t\|\leq \sqrt{\frac{\sigma^2(e^{2ct}-1)}{2c}(\varepsilon_1n+\varepsilon_2\log(1/\delta))}.
\end{equation}
This is the same, up to some constants chosen by convention, as the tight bound \eqref{eq:bound2} calculated using Gaussian concentration properties \cite[Chapter 4.4]{vershynin2018high}. For Assumption \ref{as: boundness} with time-varying $c_t$ and $\sigma_t$, we can construct linear dynamics 
    \begin{eqnarray*}
        dX_t &=& c_t X_t dt + \sigma_t dW_t
        \\
        \dot x_t &=& c_t x_t.
    \end{eqnarray*}
With the same initial condition $X_0=x_0$, $X_t-x_t$ is a zero-mean Gaussian random variance with covariance $\Psi_tI_n$ where $\Psi_t$ is as in \eqref{eq:Psi}. The Gaussianity of $X_t-x_t$ leads to the same probabilistic bound as \eqref{eq: thm1}. Therefore, Theorem \ref{thm: sto vib} is tight and cannot be improved.

While in this work the probabilistic bound in Theorem \ref{thm: sto vib} is designed for reachability analysis, we emphasize that this very bound is one of the first non-conservative results that can quantitatively describe the behavior of a stochastic system. The bound is of independent interests beyond reachability analysis and can potentially impact many other areas such as estimation, uncertainty quantification, finance, etc. 

\section{Probabilistic Reachable Set}\label{sec: PRS}
Equipped with the probabilistic bound \eqref{eq: thm1} for the stochastic deviation, we are ready to present our approach to approximating the $\delta$-PRS of a general nonlinear stochastic system \eqref{eq:stochastic}. Recalling the separation strategy in Proposition \ref{prop: separation}, we can combine our tight bound \eqref{eq: thm1} with any existing methods for approximating the DRS of the associated deterministic system \eqref{eq:associate-deterministic} to estimate the $\delta$-PRS of \eqref{eq:stochastic}.

\begin{theorem}\label{thm: PRS} 
Consider the stochastic system \eqref{eq:stochastic} with initial set $\mX_0\subseteq\R^n$ and input set $\mU\subseteq\R^p$. Suppose Assumption \ref{as: boundness} holds. Let $\overline{\mR}_t$ be an over-approximation of the DRS of the associated deterministic system \eqref{eq:associate-deterministic}. Then, for any probability level $\delta\in(0,1)$, a $\delta$-PRS of \eqref{eq:stochastic} is 
    \begin{equation}\label{eq:dPRS}
        \mR_{\delta,t}=\overline{\mR}_t \oplus \ball{r_{\delta,t},0},
    \end{equation}
    where $r_{\delta,t}=\sqrt{\Psi_t(\varepsilon_1n+\varepsilon_2\log(1/\delta))}$ with $\Psi_t$ in \eqref{eq:Psi} and $\varepsilon_1,\varepsilon_2$ in \eqref{eq: epsilon val}. 
\end{theorem}
\begin{proof}
The result follows by replacing $r_{\delta,t}$ in Proposition \ref{prop: separation} by \eqref{eq: thm1} in Theorem \ref{thm: sto vib}.
\end{proof}
Theorem \ref{thm: PRS} is a paradigm shift and essentially reduces the probabilistic reachability problem into a widely studied deterministic reachability problem. To compute the $\delta$-PRS \eqref{eq:dPRS} for the stochastic system \eqref{eq:stochastic}, one only needs to over-approximate the DRS for the deterministic system \eqref{eq:associate-deterministic}.
Theoretically speaking, the $\delta$-PRS in Theorem \ref{thm: PRS} is tight and cannot be improved further without additional assumptions. 
From a practical point of view, by combining the tight high probability bounds on stochastic deviation in Theorem~\ref{thm: sto vib} with the scalable deterministic reachability frameworks~\cite{JM-MA:15,PJM-AD-MA:19,SC:20}, the $\delta$-PRS in Theorem \ref{thm: PRS} can be computed efficiently for high-dimensional systems.

\textit{Tightness.} To be more precise, replacing $\overline{\mR}_t$ by $\mR_t$ in \eqref{eq:dPRS} gives a tight $\delta$-PRS. First, the probabilistic bound $r_{\delta,t}$ is tight provided the coefficients $c_t, \sigma_t$ in Assumption \ref{as: boundness} is tight. Moreover, since the deterministic input and stochastic disturbance in \eqref{eq:stochastic} affects the $\delta$-PRS in Definition \ref{def: p-PRS} in an independent manner, the separation strategy (Proposition \ref{prop: separation}) is also tight, meaning the decomposition in Proposition \ref{prop: separation} is necessary. Thus, the tightness of $\mR_{\delta,t}$ in Theorem \ref{thm: PRS} depends only on the tightness of the over-approximation $\overline{\mR}_t$ of the DRS of the associated deterministic system \eqref{eq:associate-deterministic}. It becomes tighter as $\overline{\mR}_t\to\mR_t$.

\textit{Computational complexity.} The computational cost of \eqref{eq:dPRS} comes from two sources: computing $\overline{\mR}_t$ and realizing the Minkowski sum $\oplus$. The former depends on the choice of algorithms for approximating DRS. Computing the Minkowski sum in a parametrized form is challenging and efficient algorithms are only available for ellipsoids and polyhedral \cite{PG-BS:93,varadhan2004accurate,weibel2007minkowski}. Fortunately, a parametrized Minkowski sum is not needed for reachability analysis. In practice, we only need an efficient membership oracle to determine whether a point $x$ belongs to the Minkowski sum, which is an easier task. In particular, for \eqref{eq:dPRS}, this oracle requires comparing $\min_{y\in\overline{\mR}_t} \|y-x\|$ and $r_{\delta,t}$, which is a convex optimization when $\overline{\mR}_t$ is convex.
In the following section, we exemplify our framework with two popular methods for computing $\overline{\mR}_t$. These methods are scalable and result in convex $\overline{\mR}_t$, rendering efficient algorithms for probabilistic reachability analysis.  
\textit{Extension to weighted norm.}
Similar to stochastic deviation, Theorem \ref{thm: PRS} is also extendable to the case for $P$-weighted $\ell_2$ norm. Consider the modified assumption as shown in Section \ref{sec:weighted}. Following the proof of Proposition \ref{prop: separation} while substituting $\mathcal{B}^n(r_{\delta,t},0)$ by $\mathcal{B}_P^n(r_{\delta,t},0)$, where $\mathcal{B}_P^n(r_{\delta,t},0)=\{x\in\R^n:~\|x\|_P\leq r_{\delta,t}\}$ is an ellipsoid, we conclude that
\begin{equation*}
    \mR_{\delta,t}=\overline{\mR}_t\oplus\mathcal{B}_P^n(r_{\delta,t},0)
\end{equation*}
is a $\delta$-PRS of the system \eqref{eq:stochastic}. 

\section{Case study of Probabilistic Reachability}\label{sec: app}
In this section, we present the application of $\delta$-PRS derived in Section \ref{sec: PRS} in two case studies where contraction-based and interval-based methods are used to approximate $\mathcal{R}_t$.

\subsection{Contraction-based Probabilistic Reachability}
Contraction theory is a classical framework for analyzing the stability of dynamical systems using the incremental distance between their trajectories~\cite{WL-JJES:98,ZA-EDS:14}. Traditionally, it is employed to infer strong robustness properties of dynamical systems. 
Recently, contraction theory has emerged as a computationally efficient tool for reachability analysis of deterministic systems. 
The contraction-based method relies on the matrix measure (Definition \ref{def:matrix}) and the following assumption.
\begin{assumption}\label{assum:2}
 For the deterministic system~\eqref{eq:associate-deterministic}, there exist constants $c,\rho\in \real$ such that, for every $t,x,u\in \real_{\ge 0}\times \real^n\times \mathcal{U}$,  
 \begin{enumerate}
     \item $\mu_{\mathbb{X}}(D_xf(x,u,t))\le c$, and
     \item $\|D_u f(x,u,t)\|_{\mathbb{X},\mathbb{U}}\le \rho$.
 \end{enumerate} 
\end{assumption}
Here $\mu_{\mathbb{X}}$ is the matrix measure with respect to the norm $\|\cdot\|_{\mathbb{X}}$ on $\real^n$ and $\|\cdot\|_{\mathbb{X},\mathbb{U}}$ denotes the induced norm on $\real^{p\times n}$. The norm $\|\cdot\|_{\mathbb{X}}$ can be chosen differently from the Euclidean norm in general to ensure the tightest possible reachable set. 
Suppose that system~\eqref{eq:associate-deterministic} satisfies Assumption~\ref{assum:2} and let $t\mapsto x^*_t$ be a trajectory of~\eqref{eq:associate-deterministic} with the input $t\mapsto u^*_t$. Given initial configuration $x_0\in\mathcal{X}_0 = \mathcal{B}_{{\mathbb{X}}}(r_1,x^*_0)$ for $r_1>0$ and input $u_t\in\mathcal{B}_{{\mathbb{U}}}(r_2,u^*_t)\subset\mathcal{U}$ for $r_2>0$, the contraction-based method gives the following over-approximation of reachable sets of~\eqref{eq:associate-deterministic}~\cite{AD-SJ-FB:20o} 
\begin{equation}\label{eq:approx-contraction}
    \overline{\mathcal{R}}_t= \mathcal{B}_{{\mathbb{X}}}(e^{ct}r_1+ \tfrac{\rho}{c} (e^{ct}-1)r_2 , x^*_t). 
\end{equation}
The contraction-based over-approximation of reachable sets in~\eqref{eq:approx-contraction} can be combined with Theorem~\ref{thm: PRS} to estimate a $\delta$-PRS of the system~\eqref{eq:stochastic}.

\begin{proposition}[Contraction-based reachability]\label{prop:cbr}
Consider the stochastic system~\eqref{eq:stochastic} and its associated deterministic system~\eqref{eq:associate-deterministic} satisfying Assumptions~\ref{as: boundness} and~\ref{assum:2}. Let $t\mapsto x^*_t$ be a trajectory of~\eqref{eq:associate-deterministic} with the input $t\mapsto u^*_t$ and $t\mapsto X_t$ be a trajectory of the stochastic system~\eqref{eq:stochastic} starting from $x_0\in \mathcal{B}_{{\mathbb{X}}}(r_1,x^*_0)$ with an input $u_t:\real_{\ge 0}\to \mathcal{B}_{{\mathbb{U}}}(r_2,u^*_t)$. Then, for every $t\ge 0$, with probability at least $1-\delta$, 
\begin{align*}
    X_t \in \mathcal{B}_{{\mathbb{X}}}(e^{ct}r_1+ \tfrac{\rho}{c} (e^{ct}-1)r_2 , x^*_t) \oplus \mathcal{B}^n(r_{\delta,t},0)
\end{align*}
where $r_{\delta,t}=\sqrt{\Psi_t(\varepsilon_1n+\varepsilon_2\log(1/\delta))}$, $\Psi_t$ is as in \eqref{eq:Psi}, and $\varepsilon_1$,$\varepsilon_2$ are given by \eqref{eq: epsilon val}.
\end{proposition}
\begin{proof}
    The result follows by combining Theorem~\ref{thm: PRS} and the contraction-based over-approximation of the reachable set of system~\eqref{eq:associate-deterministic} in~\eqref{eq:approx-contraction}. 
\end{proof}

\subsection{Interval-based Probabilistic Reachability}

 Interval analysis is a framework for estimating propagation of uncertainties by computing function bounds~\cite{LJ-MK-OD-EW:01} and has been successfully used for reachability analysis of deterministic systems.   
 The main idea of interval-based reachability is to embed the dynamical system into a higher dimensional space using a suitable inclusion function. 
 The map $\left[\begin{smallmatrix}\underline{\OF}\\ \overline{\OF}\end{smallmatrix}\right]: \real^{2n}\times \real^{2p}\times \R_{\ge 0}\to \real^{2n}$ is an inclusion function for $f$, if, for every $z,w\in [\underline{x},\overline{x}]\times [\underline{u},\overline{u}]$ and every $t\ge 0$, 
 \begin{align*}
     \underline{\OF}(\underline{x},\overline{x},\underline{u},\overline{u},t) \le f(z,w,t) \le \overline{\OF}(\underline{x},\overline{x},\underline{u},\overline{u},t). 
 \end{align*}
 Many automated approaches exist for finding an inclusion function for $f$. We refer to~\cite[Section IV.B]{SJ-AH-SC:23} for a detailed discussion on these approaches and to~\cite{harapanahalli2023toolbox} for a toolbox for computing inclusion functions.  
 
 Given an interval initial configuration $\mathcal{X}_0=[\underline{x}_0,\overline{x}_0]$ and an interval input set $\mathcal{U}=[\underline{u},\overline{u}]$, the embedding system of~\eqref{eq:associate-deterministic} associated with the inclusion function $\OF$ is given by
\begin{align}\label{eq:embedding}
  \begin{bmatrix}
      \dot{\underline{x}}\\
      \dot{\overline{x}}
  \end{bmatrix} = \begin{bmatrix}
      \underline{\OF}(\underline{x},\overline{x},\underline{u},\overline{u},t)\\
      \overline{\OF}(\underline{x},\overline{x},\underline{u},\overline{u},t)
  \end{bmatrix}.  
\end{align}
Let $\left[\begin{smallmatrix}\underline{x}_t\\ \overline{x}_t\end{smallmatrix}\right]$ be the trajectory of the embedding system~\eqref{eq:embedding} starting from  $\left[\begin{smallmatrix}
    \underline{x}_0\\ \overline{x}_0
\end{smallmatrix}\right]$. Then, an over-approximation of the deterministic reachable set of~\eqref{eq:associate-deterministic} is~\cite[Proposition 5]{SJ-AH-SC:23}
\begin{align}\label{eq:interval}
    \overline{\mathcal{R}}_t= [\underline{x}_t,\overline{x}_t].
\end{align}
This interval-based over-approximation of reachable sets can be combined with Theorem~\ref{thm: PRS} to estimate a $\delta$-PRS of the system~\eqref{eq:stochastic}.

\begin{proposition}[Interval-based reachability]\label{prop:mbr}
Consider the stochastic system \eqref{eq:stochastic} and its associated deterministic system \eqref{eq:associate-deterministic} satisfying Assumption~\ref{as: boundness}. 
Let $t\mapsto X_t$ be a trajectory of the stochastic system~\eqref{eq:stochastic} starting from $x_0\in [\underline{x}_0,\overline{x}_0]$ with an input curve $u_t:\real_{\ge 0}\to [\underline{u},\overline{u}]$. Suppose that $\OF = \left[\begin{smallmatrix}\underline{\OF}\\ \overline{\OF}\end{smallmatrix}\right]$ is an inclusion function for $f$ and $\left[\begin{smallmatrix}\underline{x}_t\\ \overline{x}_t\end{smallmatrix}\right]$ is the trajectory of the embedding system~\eqref{eq:embedding} starting from $\left[\begin{smallmatrix}
    \underline{x}_0\\ \overline{x}_0
\end{smallmatrix}\right]$. Then, for every $t\ge 0$, with probability at least $1-\delta$
\begin{align*}
    X_t \in [\underline{x}_t,\overline{x}_t]\oplus\mathcal{B}^n(r_{\delta,t},0),
\end{align*}
where $r_{\delta,t}=\sqrt{\Psi_t(\varepsilon_1n+\varepsilon_2\log(1/\delta))}$, $\Psi_t$ is as in \eqref{eq:Psi}, $\varepsilon_1$,$\varepsilon_2$ are given by \eqref{eq: epsilon val}. 
\end{proposition}
\begin{proof}
    The result follows by combining Theorem~\ref{thm: PRS} and the interval over-approximation of the reachable set of the deterministic system~\eqref{eq:associate-deterministic} in \eqref{eq:interval}. 
\end{proof}

\begin{figure}[t]
	\centering
        \begin{subfigure}[t]{0.24\textwidth}
			\centering
			\includegraphics[width=1.0\textwidth]{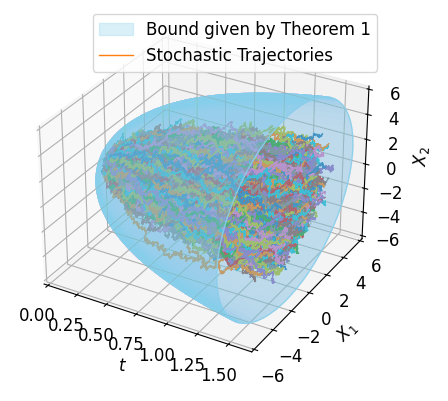}
			\caption{Trajectories of $X_t$}
	\end{subfigure}%
        \begin{subfigure}[t]{0.24\textwidth}
		\centering
		\includegraphics[width=1.1\textwidth]{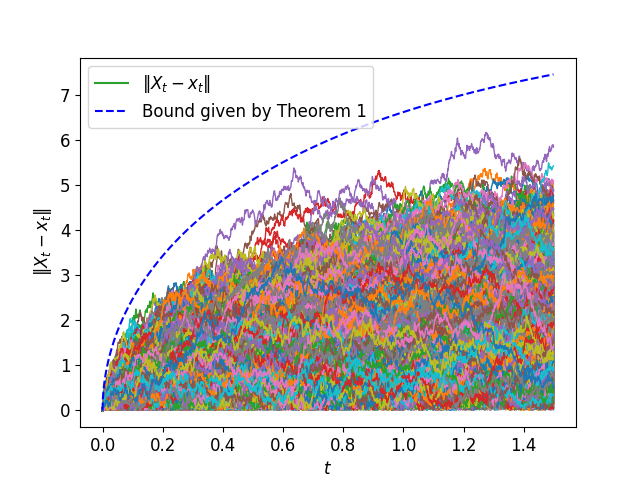}
		\caption{Trajectories of $\|X_t-x_t\|$}
	\end{subfigure}%
	\caption{Probabilistic bound of stochastic deviation for a linear system. In Figure \ref{fig: Lin set}(a), each curve represents an independent trajectory of $X_t$. In Figure \ref{fig: Lin set}(b), each solid curve is an independent trajectory of $\|X_t-x_t\|$. The blue envelope and the blue dashed curve correspond to our bound \eqref{eq: thm1}. 
 }
	\label{fig: Lin set}
\end{figure} 

\begin{figure}[t]
	\centering
        \begin{subfigure}[t]{0.24\textwidth}
			\centering
			\includegraphics[width=1.1\textwidth]{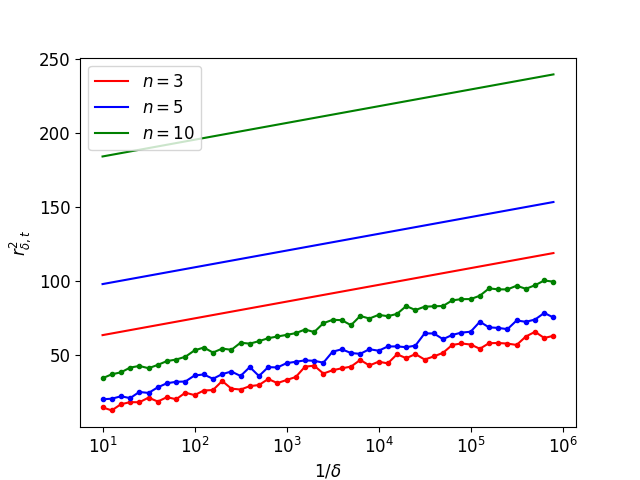}
			\caption{$r_{\delta,t}^2$ w.r.t. $1/\delta$}
	\end{subfigure}%
        \begin{subfigure}[t]{0.24\textwidth}
		\centering
		\includegraphics[width=1.1\textwidth]{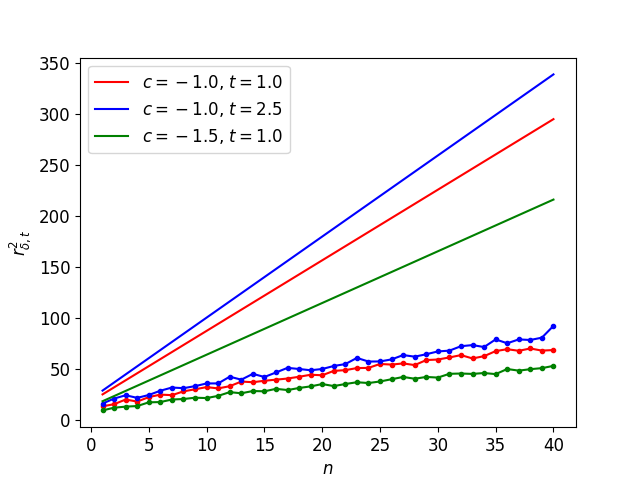}
		\caption{$r_{\delta,t}^2$ w.r.t. $n$}
	\end{subfigure}%
	\caption{Illustration of the tightness of $r_{\delta,t}$ w.r.t. $\delta,n$. In Figure \ref{fig: r-delta r-n}(a), the solid line shows the dependence of $r_{\delta,t}^2$ over $1/\delta$ and the dotted line in the same color is the corresponding simulated bound $\hat{r}_{\delta,t}^2$. In Figure \ref{fig: r-delta r-n}(b), the solid line shows the dependence of $r_{\delta,t}^2$ over $n$, and the dotted line in the same color is the corresponding simulated $\hat{r}_{\delta,t}^2$.  
 }
	\label{fig: r-delta r-n}
\end{figure}

\section{Numerical experiments}\label{sec: simulations}

In this section, we present several examples to illustrate the efficacy of our framework and the tightness of our results.

\subsection{Linear Example}

We first consider a linear example to validate the tightness of our bound \eqref{eq: thm1} on the stochastic deviation. Consider a simple linear dynamics

\begin{equation}\label{sys: linear}
    \begin{split}
        dX_t&=-0.4I_nX_tdt+ \sqrt{2}dW_t\\
        &= AX_tdt+\sigma dW_t,
    \end{split}
\end{equation}
initialized at $X_0=0$. The system \eqref{sys: linear} satisfies Assumption \ref{as: boundness} with $c_t\equiv c = \mu(A)=-0.4$ and $\sigma_t\equiv\sigma=\sqrt{2}$. By linearity, $X_t$ follows a zero-mean Gaussian distribution whose covariance $\cov{X_t}$ can be computed using \eqref{eq: cov(X_t)} in closed-form. The trajectory of the deterministic dynamics associated with \eqref{sys: linear} starting from $x_0=0$ is $x_t\equiv 0$.

To illustrate the bound \eqref{eq: thm1}, we simulate 5000 independent trajectories of \eqref{sys: linear} with $n=2$ over a time horizon $t\in[0,1.5]$ and compute the deviation associated with each trajectory, as depicted in Figure \ref{fig: Lin set}. These trajectories are compared with our probabilistic bound $r_{\delta,t}$ with design parameter $\varepsilon=1/16$, $\delta=10^{-3}$. Figure \ref{fig: Lin set} shows that all the trajectories satisfy the bound $r_{\delta,t}$ as expected.

By Theorem \ref{thm: sto vib}, the square of our bound \eqref{eq: thm1}, $r_{\delta,t}^2$, grows linearly with $\log(1/\delta)$ and $n$, as illustrated in Figure \ref{fig: r-delta r-n}. To verify the tightness of these dependencies, we compare them with those obtained through simulation. In particular, for each choice of $\delta$ and $n$, we simulate $10^7$ independent trajectories of \eqref{sys: linear} and compute the associated value of $\|X_t-x_t\|$ for each trajectory. We follow a standard approach \cite{shapiro2003monte} and estimate the high probability bound $\hat{r}_{\delta,t}$ of the stochastic deviation as the $\delta$-th largest $\|X_t-x_t\|$ (e.g., top 1\% if $\delta=10^{-2}$). The results, shown in Figure \ref{fig: r-delta r-n}, imply that $\hat{r}_{\delta,t}^2$ also grows linearly with $\log(1/\delta)$ and $n$, consistent with our theoretical bound \eqref{eq: thm1}. 

Note that there is a gap between the calculated bounds with $\varepsilon=1/16$ and the simulated bounds in \ref{fig: r-delta r-n}. This is due to the choice of parameters $\varepsilon_1$ and $\varepsilon_2$. These parameters $\varepsilon_1$ and $\varepsilon_2$ \eqref{eq: epsilon val} are constructed in the proof for all $\delta, n$ and are not optimal for each choice of $\delta, n$, as explained in Section \ref{subsec: motivation}. 

\subsection{Inverted Pendulum}\label{subsec: SP}
Next, we consider an inverted pendulum with a stabilizing state feedback controller, whose state space model is given by 
\begin{equation}\label{sys: SP}
    \begin{split}
        dX_t&= \begin{bmatrix}
            \dot{\theta} \\
            \frac{g}{L}\sin\theta+ KX_t \\
        \end{bmatrix}dt+g_tdW_t \\
    \end{split}
\end{equation}
where $X_t=\begin{bmatrix} \theta&\dot{\theta} \end{bmatrix}^{\top}$ is the state vector, $\theta$ is the angle describing the position of the pendulum, $\dot{\theta}$ is the angular velocity of the pendulum, $KX_t = K_1\theta + K_2\dot{\theta}$ is a stabilizing linear state feedback controller, and $g_tdW_t$ is the stochastic disturbance on the angular acceleration with $W_t$ a one-dimensional Wiener process. Set the gravity $g=10$, the pendulum length $L=1$, and $g_t =\begin{bmatrix}0 & 0.1\end{bmatrix}^{\top}$. 
The linear state feedback controller $KX_t$ is designed with feedback gain $K=\begin{bmatrix} K_1 & K_2\end{bmatrix}=\begin{bmatrix} -20 & -20 \end{bmatrix}$ to stabilize the equilibrium point $x^*=\begin{bmatrix}\theta^* & \dot{\theta}^*\end{bmatrix}^{\top} = 0$ of the associated deterministic system 
\begin{align}\label{eq:ip-determine}
 \dot{x}_t =\begin{bmatrix}\dot{\theta} \\ 
 \frac{g}{L}\sin(\theta)+ K_1\theta + K_2\dot{\theta}\end{bmatrix} : = f(x_t),   
\end{align}
 where $x_t = \begin{bmatrix}\theta & \dot{\theta}\end{bmatrix}^{\top}$.

Our goal is to find a tight $\delta$-PRS of the inverted pendulum~\eqref{sys: SP} starting from the initial configuration $\mathcal{X}_0=[-\frac{\pi}{10},\frac{\pi}{10}]\times [-0.2,0.2]$.   
We use Theorem~\ref{thm: PRS} with contraction-based and interval-based deterministic reachability methods to obtain $\delta$-PRS of the inverted pendulum~\eqref{sys: SP}. 
We first consider the modified version of Assumption~\ref{as: boundness} introduced in Section~\ref{sec:weighted} as $\mu_{P}(D_x f(x))\le c_t$ and $P^{\frac{1}{2}}g_t g_t^{\top}P^{\frac{1}{2}} \preceq \sigma_t^2 I_n$ for every $t\ge 0$ and $x\in \real^n$. 
For every $x = (\theta, \dot{\theta})^{\top}\in \real^2$,
\begin{align*}
    D_xf(x) = \left[\begin{smallmatrix}  0 & 1 \\ \frac{g}{L}\cos(\theta)+K_1 & K_2\end{smallmatrix}\right].
\end{align*}
 We define the matrices $A_1,A_2\in \real^{2\times 2}$ as follows:
 \begin{align*}
  A_1=\left[\begin{smallmatrix}  0 & 1 \\ \frac{g}{L}+K_1 & K_2\end{smallmatrix}\right], \qquad A_2=\left[\begin{smallmatrix}  0 & 1 \\ -\frac{g}{L}+K_1 & K_2\end{smallmatrix}\right].  
 \end{align*}
 Note that $\cos(\theta)\in [-1,1]$. This implies that, for every $x\in \real^2$, we have $D_x f(x) \in \mathrm{conv}\left\{A_1, A_2\right\}$, where $\mathrm{conv}$ is the convex hull. Thus, using~\cite[Lemma 4.1]{CF-JK-XJ-SM:18}, the minimum constant contraction rate $c_t=c$ for the system~\eqref{eq:ip-determine} can be computed using the following optimization problem:
\begin{align}\label{eq:optim}
   \min_{c\in \real, P \succ 0}&\quad c\nonumber\\
   &\mbox{s.t.}\;\; A_i^{\top} P + P A_i \preceq 2c P,\quad\mbox{for }i\in \{1,2\}.
\end{align}
We solve optimization problem~\eqref{eq:optim} by successively applying semi-definite programming on $P$ and bisection on $c$. The optimal solution of~\eqref{eq:optim} is given by the constant contraction rate $c_t=c=-0.5$ and the weight matrix $P=\left[\begin{smallmatrix} 35.68  &  2.21\\
    2.21  &  1.27
\end{smallmatrix}\right]$. 
With this matrix $P$, we compute $P^{\frac{1}{2}}g_t g_t^{\top}P^{\frac{1}{2}} = \left[\begin{smallmatrix}0.0010 & 0.0034\\
    0.0034 & 0.0118\end{smallmatrix}\right]\preceq 0.0128 I_2$,
and get $\sigma_t=\sigma = 0.1130$.

 \begin{figure}[ht]
 \centering
  \includegraphics[width =0.48\linewidth]{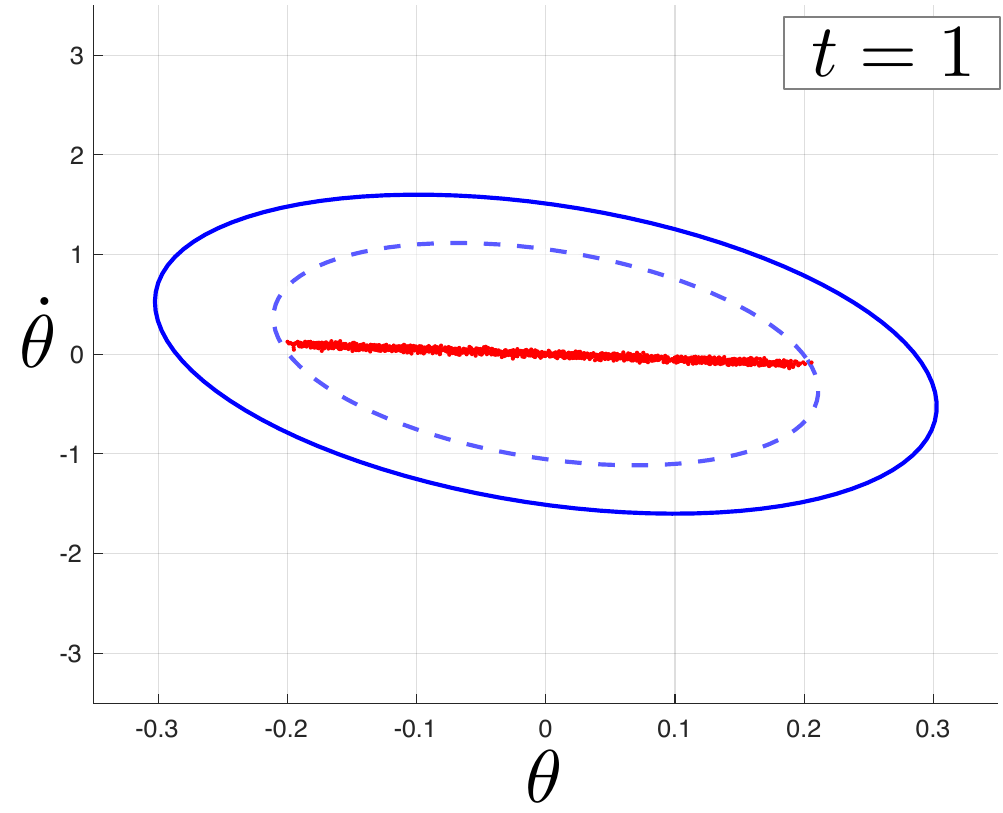}
  \includegraphics[width =0.48\linewidth]{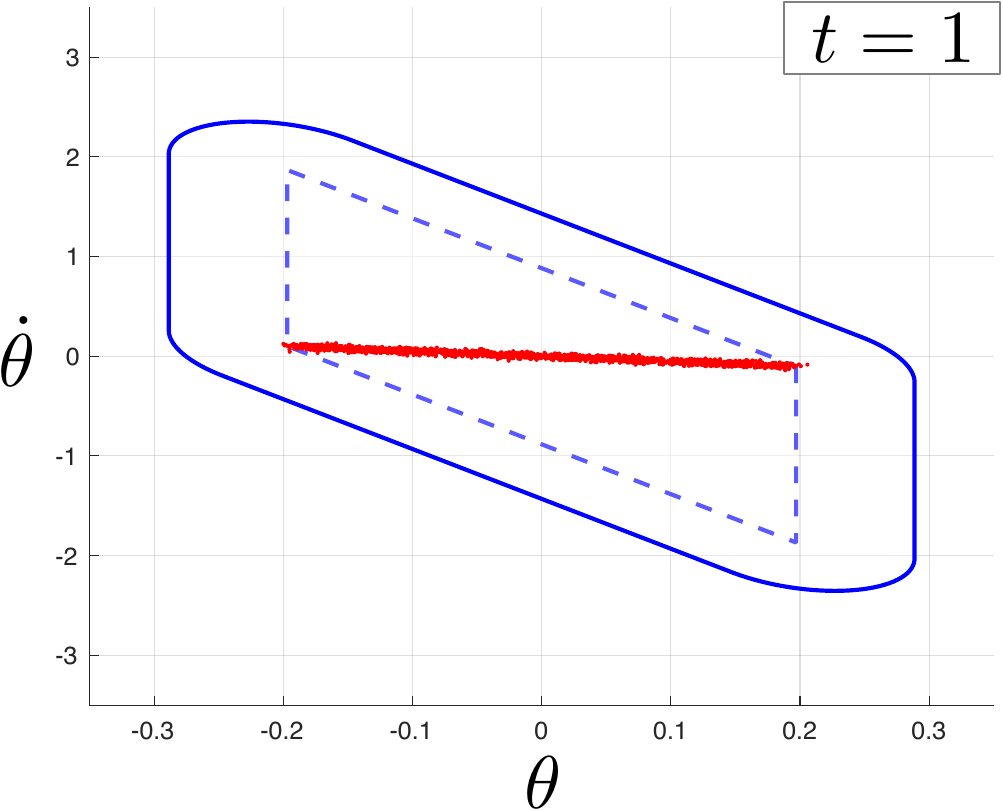}
  \includegraphics[width =0.48\linewidth]{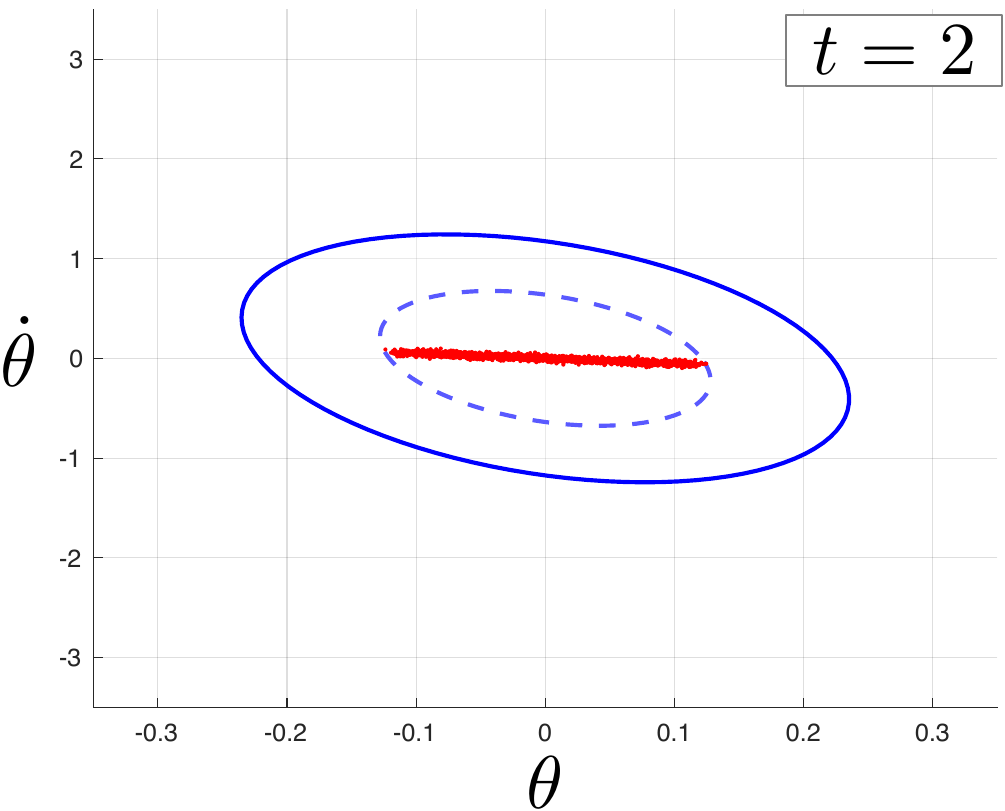}
  \includegraphics[width =0.48\linewidth]{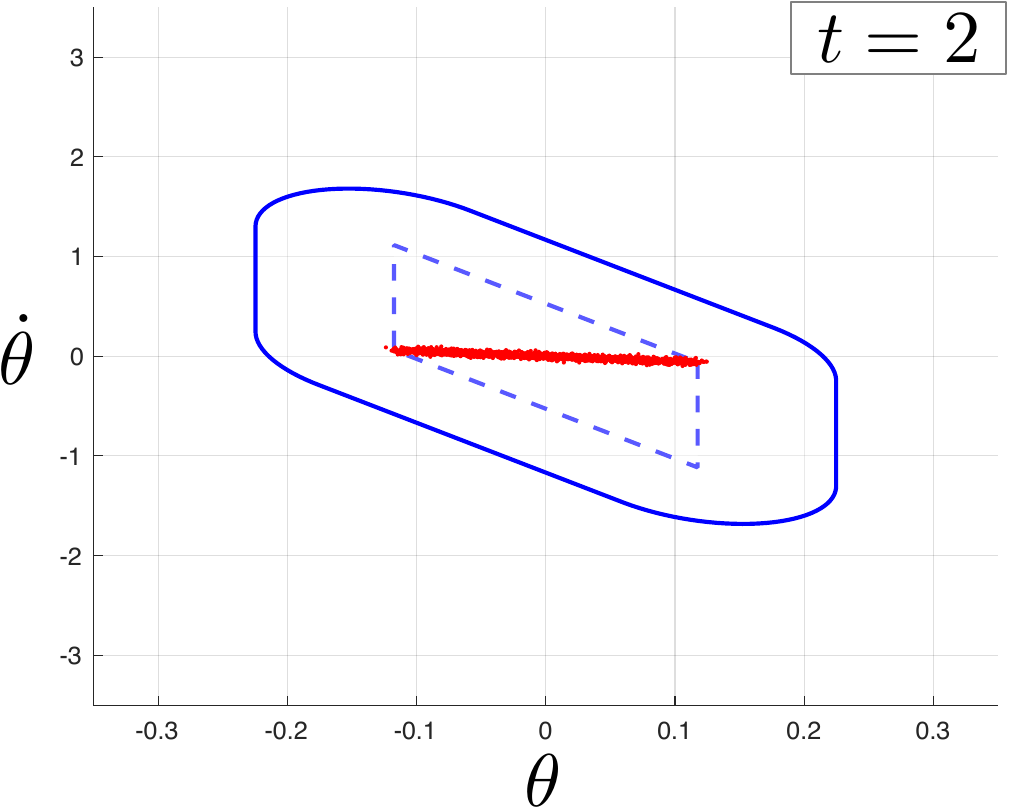}
  \includegraphics[width =0.48\linewidth]{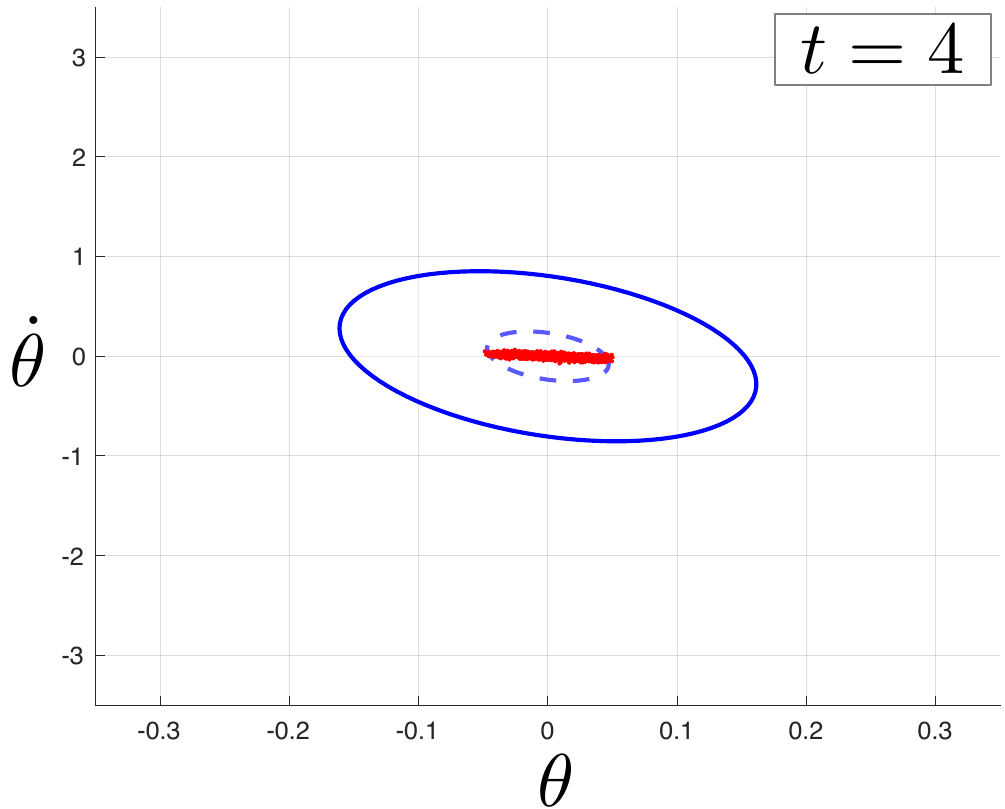}
  \includegraphics[width =0.48\linewidth]{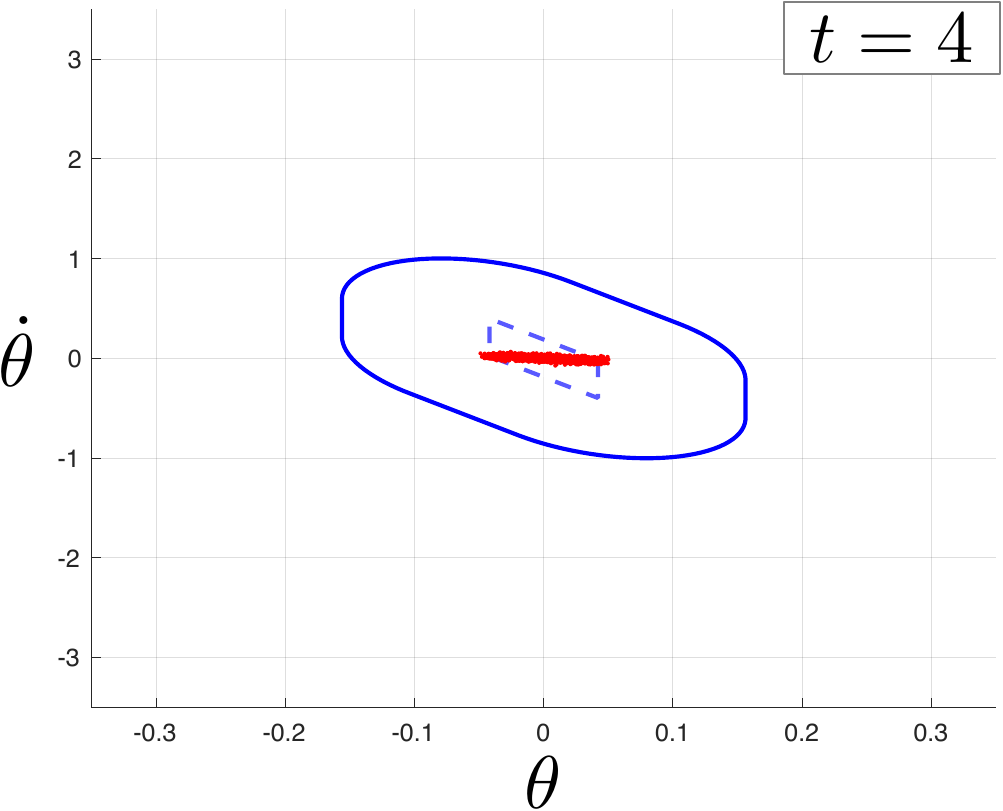}
  \caption{\textbf{Left}: The solid blue lines show the boundary of $\delta$-PRS with $\delta=10^{-3}$ at times $t=1,2,4$ for the stochastic inverted pendulum~\eqref{sys: SP} starting from $\overline{\mathcal{X}_0}\supset \mathcal{X}_0$ obtained using Proposition~\ref{prop:cbr}. The dashed blue lines are the boundary of the ellipsoids that over-approximate reachable sets of the associated deterministic system~\eqref{eq:ip-determine}. The red dots are $2000$ random trajectories of the inverted pendulum~\eqref{sys: SP} starting from $\overline{\mathcal{X}}_0\supset \mathcal{X}_0$ at times $t=1,2,4$.
  \textbf{Right}: The solid blue lines show the boundary of $\delta$-PRS with $\delta=10^{-3}$ at times $t=1,2,4$ for the inverted pendulum~\eqref{sys: SP} starting from $T^{-1}\overline{\mathcal{Y}}_0\supset \mathcal{X}_0$ obtained using Theorem~\ref{thm: PRS} and interval-based reachability of the transformed system. The dashed blue lines are the boundary of the parallelotopes obtained from the interval analysis that over-approximates the reachable sets of the associated deterministic system~\eqref{eq:ip-determine}. The red dots are $2000$ random trajectories of the inverted pendulum~\eqref{sys: SP} starting from $T^{-1}\overline{\mathcal{Y}}_0\supset \mathcal{X}_0$ at times $t=1,2,4$.}
	\label{fig:pendulum}
 \end{figure}

\paragraph*{Contraction-based Reachability} 
We use Proposition~\ref{prop:cbr} to find a $\delta$-PRS of~\eqref{sys: SP}. 
We consider Assumption~\ref{assum:2} with $\|\cdot\|_{\mathbb{X}}=\|\cdot\|_{P}$ with positive definite matrix $P$ as defined above. For every $x \in \real^2$, we have $\mu_{P}(D_xf(x))\le c = -0.5$. Using Proposition~\ref{prop:cbr} with the initial configuration $\overline{\mathcal{X}}_0=\setdef{x\in \real^2}{\|x\|_{P}\le \left\|\left[\begin{smallmatrix}\frac{\pi}{10}\\  0.2\end{smallmatrix}\right]\right\|_{P}}\supset \mathcal{X}_0$, we obtain a $\delta$-PRS of~\eqref{sys: SP} with $\delta = 10^{-3}$ as shown in Figure~\ref{fig:pendulum} (left). 

\paragraph*{Interval-based Reachability} We use Theorem~\ref{thm: PRS} with a modified version of interval-based analysis for the associated deterministic system~\eqref{eq:ip-determine} to find a $\delta$-PRS of~\eqref{sys: SP}. 
We consider the coordinate transformation $y_t=Tx_t$ with nonsingular matrix $T=\left[\begin{smallmatrix}1 & 0.2\\ 1 & 0\end{smallmatrix}\right]$ for the associated deterministic system~\eqref{eq:ip-determine} and apply interval-based reachability to the transformed system. 
We employ Theorem~\ref{thm: PRS} with the initial configuration $T^{-1}\overline{\mathcal{Y}}_0\supset \mathcal{X}_0$ where $\overline{\mathcal{Y}}_0 = [{-\tfrac{\pi}{10}\left[\begin{smallmatrix}1.04\\ 1\end{smallmatrix}\right]},{\tfrac{\pi}{10}\left[\begin{smallmatrix}1.04\\ 1\end{smallmatrix}\right]}]$.
The $\delta$-PRS of~\eqref{sys: SP} with $\delta = 10^{-3}$ obtained using this analysis are shown in Figure~\ref{fig:pendulum} (right).

\subsection{Nonlinear Unicycle model}
\begin{figure}
  \includegraphics[width =0.49\linewidth]{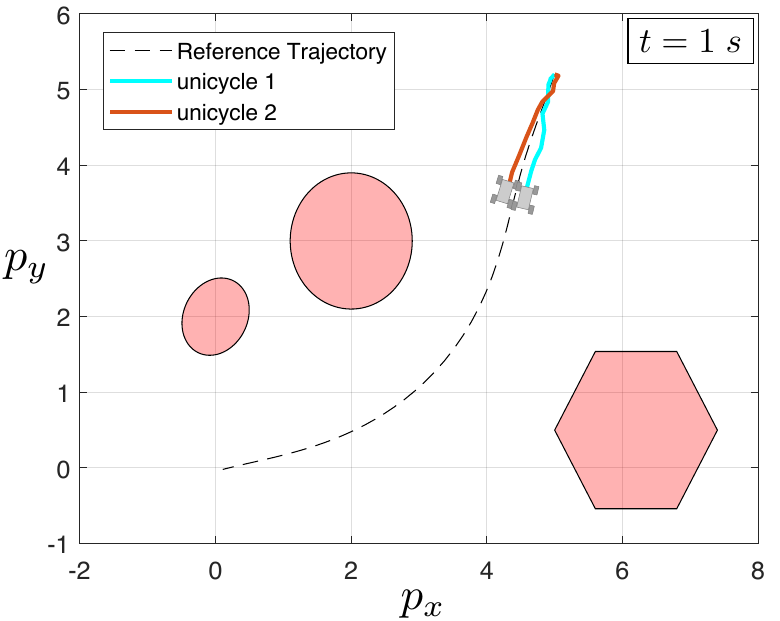}
  \includegraphics[width =0.49\linewidth]{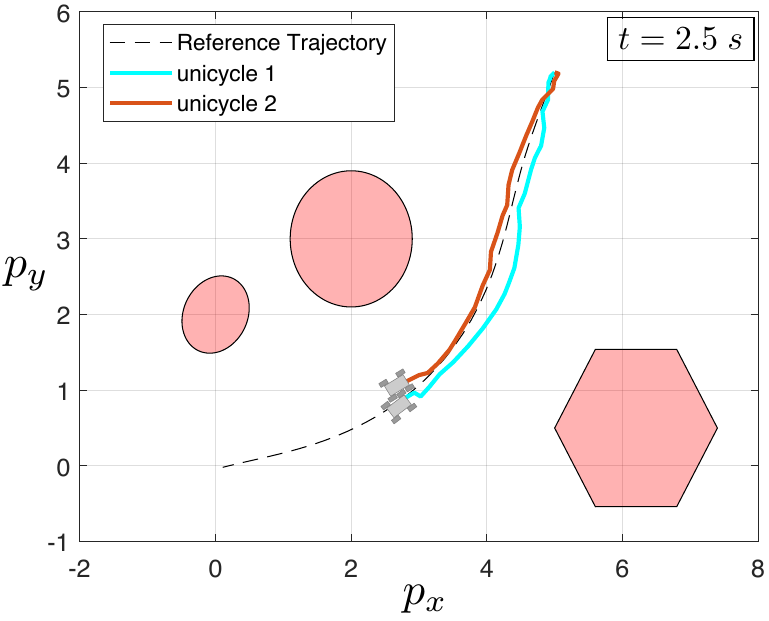}
  \includegraphics[width =0.49\linewidth]{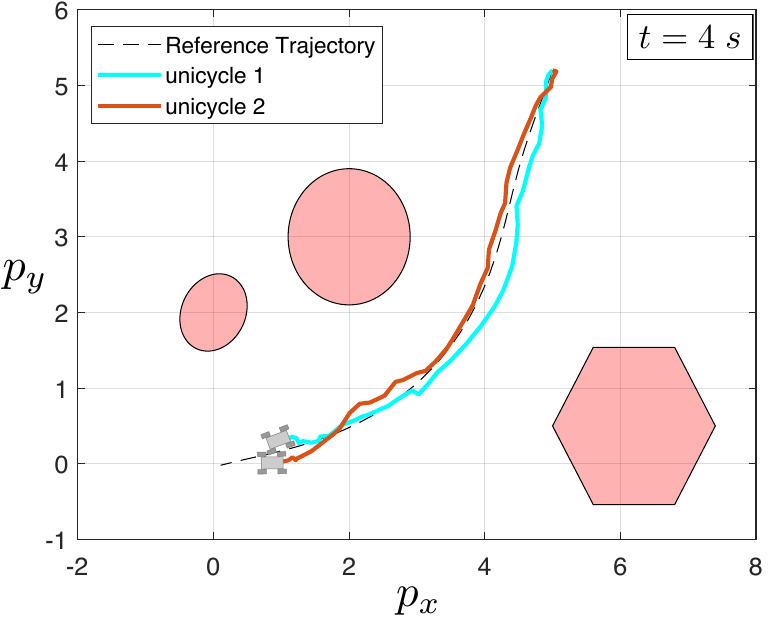}
  \includegraphics[width =0.49\linewidth]{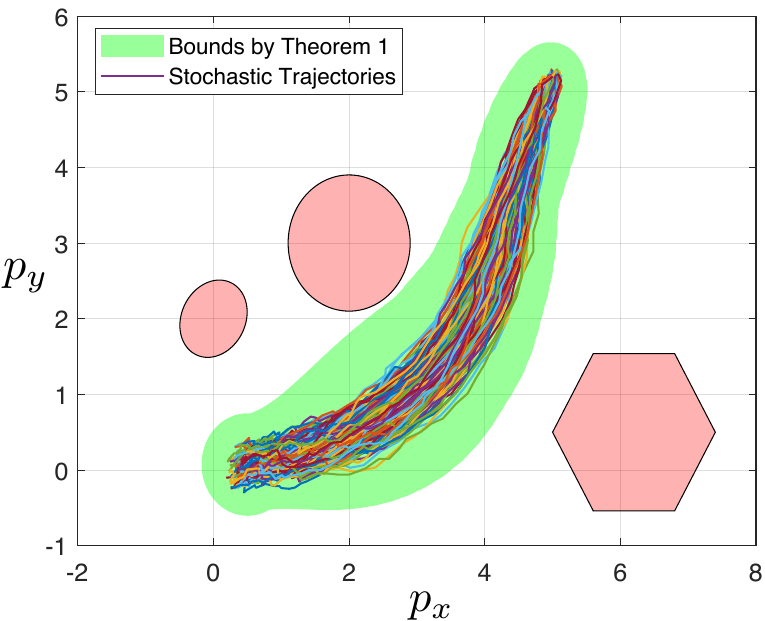}
  
	\caption{The stochastic vehicle~\eqref{eq:unicycle} with the tracking feedback controller~\eqref{eq:polar-controller-unicycle} in the $p_x-p_y$ plane with obstacles shown in light red. The reference trajectory designed using MPC for the deterministic system~\eqref{eq:unicycle-d} is shown with dashed black line. \textbf{Top left}: Two stochastic trajectories of vehicle~\eqref{eq:unicycle} with the tracking feedback controller~\eqref{eq:polar-controller-unicycle} starting from $x_0=(5,5,-\frac{2\pi}{3})$ at time $t=1$. \textbf{Top right}: The same two stochastic trajectories at time $t=2.5$. \textbf{Bottom left}: The same two stochastic trajectories at time $t=4$. \textbf{Bottom right}: $200$ sample trajectories of the stochastic  vehicle~\eqref{eq:unicycle} with the tracking feedback controller~\eqref{eq:polar-controller-unicycle}. The green envelope is the union of $\delta$-PRS at times $t\in [0,5]$ obtained using a modified version of Proposition~\ref{prop:cbr} with $\delta = 10^{-3}$. Therefore, with probability at least $99.9 \%$, the stochastic vehicle~\eqref{eq:unicycle} with the tracking feedback controller~\eqref{eq:polar-controller-unicycle} is safe for all $t\in [0,5]$.}
	\label{fig:unicycle}
\end{figure}

Finally, we consider a vehicle moving on a $2$-dimensional plane with obstacles shown in light red in Figure~\ref{fig:unicycle}. The vehicle is  modeled by the unicycle dynamics
\begin{align}\label{eq:unicycle}
dX_t = \begin{bmatrix}
    v_t \cos(\theta)\\
    v_t \sin(\theta)\\
    w_t + u_t
\end{bmatrix} dt + g_t dW_t
\end{align}
where $X_t = \begin{bmatrix}p_x & p_y & \theta\end{bmatrix}^{\top}$ is the state of the vehicle, $(p_x,p_y)$ is the position of the center of mass of the vehicle in the plane, $\theta$ is the heading angle of the vehicle, $v_t$ is the linear velocity of the center of mass,  $w_t$ is the angular velocity of the vehicle, $u_t$ is the deterministic disturbance on the angular velocity, and $g_tdW_t$ is the stochastic disturbance on the model with $W_t$ a three-dimensional Wiener process. The associated deterministic unicycle model is given by
\begin{align}\label{eq:unicycle-d}
   \dot{x}_t = \begin{bmatrix}v_t \cos(\theta)\\ v_t \sin(\theta)\\ w_t \end{bmatrix} + \begin{bmatrix}0\\0\\ u_t\end{bmatrix}:=f(x_t,u_t,t)
\end{align}
where $x_t=\begin{bmatrix}p_x & p_y & \theta\end{bmatrix}^{\top}$. 
We use Model Predictive Control (MPC) to design an open-loop controller
to steer the deterministic system~\eqref{eq:unicycle-d} from the initial configuration $x_0=(5,5,-\frac{2\pi}{3})$ to the origin while avoiding the obstacles in the $p_x-p_y$ plane.
The trajectory of the deterministic system~\eqref{eq:unicycle-d} with the MPC controller starting from $x_0=(5,5,-\frac{2\pi}{3})$ is denoted by $t\mapsto (p_x^*,p_y^*,\theta^*)$. We consider $t\mapsto (p_x^*,p_y^*,\theta^*)$ as the \emph{reference trajectory} for the stochastic vehicle~\eqref{eq:unicycle}. 
Using the approach in~\cite{MA-GC-AB-AB:95}, we design the following feedback controller for tracking the reference trajectory $t\mapsto (p_x^*,p_y^*,\theta^*)$: 
\begin{align}\label{eq:polar-controller-unicycle}
    v_t &= K_r r_t\cos(\alpha_t),\nonumber\\
    w_t &= K_{\alpha} \alpha_t + K_r \sin(\alpha_t)\cos(\alpha_t)\tfrac{\alpha_t + \beta_t }{\alpha_t},
\end{align}
where the variables $r_t,\alpha_t,\beta_t$ are defined as
\begin{align*}
    r_t &=\sqrt{(p_x-p_x^*)^2 + (p_y -p_y^*)^2},\\
    \alpha_t &= \theta - \mathrm{atan}(p_y -p_y^*,p_x-p_x^*),\\
    \beta_t &= \mathrm{atan}(p_y-p_y^*,p_x-p_x^*)-\theta^*,
\end{align*}
and $K_r,K_{\alpha}\ge 0$ are feedback gains. 

We consider the stochastic vehicle~\eqref{eq:unicycle} with $g_t= 0.1$ and $u_t\in [-0.03,0.03]$ with the tracking controller~\eqref{eq:polar-controller-unicycle} and feedback gains $K_r = -0.8$ and $K_{\alpha} = -1.5$. 
We assume that this stochastic vehicle starts from $x_0=(5,5,-\frac{2\pi}{3})$. Our goal is to provide high probability guarantees that the stochastic vehicle~\eqref{eq:unicycle} with the tracking controller~\eqref{eq:polar-controller-unicycle} avoids the obstacles shown in Figure~\ref{fig:unicycle}, over the time horizon $[0,5]$. 
We use a modified version of Proposition~\ref{prop:cbr} to construct $\delta$-PRS of the stochastic vehicle~\eqref{eq:unicycle} with the tracking controller~\eqref{eq:polar-controller-unicycle}. 
We use the strategy in \cite{chuchu2017simulation} to estimate a time-varying $c_t$ in Assumption~\ref{as: boundness}. 
We also use a generalization of Assumption~\ref{assum:2} with $\|\cdot\|_{\mathbb{X}}$ and $\|\cdot\|_{\mathbb{U}}$ defined as standard Euclidean norms and with time-varying contraction rate $c_t$. This time-varying contraction rate is then used for contraction-based reachability analysis of the associated deterministic system~\eqref{eq:unicycle-d} in Proposition~\ref{prop:cbr}. 
For $\delta=10^{-3}$, the $\delta$-PRS of the stochastic vehicle~\eqref{eq:unicycle} with the tracking controller~\eqref{eq:polar-controller-unicycle} starting from $x_0=(5,5,-\frac{2\pi}{3})$ at times $t\in [0,5]$ are shown in Figure~\ref{fig:unicycle} using the green envelope. 
From Figure~\ref{fig:unicycle}, it is clear that the green envelope does not intersect any of the obstacles in the $p_x-p_y$ plane. 
Therefore, with probability at least $99.9 \%$, the stochastic vehicle~\eqref{eq:unicycle} with the tracking controller~\eqref{eq:polar-controller-unicycle} starting from $x_0=(5,5,-\frac{2\pi}{3})$ is safe and does not hit any obstacle for all times $t\in [0,5]$.

\section{Conclusion} \label{sec: conclusion}
We propose an efficient and flexible framework for computing the Probabilistic Reachable Set (PRS) of continuous-time nonlinear stochastic systems. 
Using a suitable separation strategy, we decouple the effect of deterministic inputs and the effect of stochastic uncertainties on the PRS.   
This separation strategy is flexible as it allows using any deterministic reachability method to capture the effects of deterministic inputs. It essentially reduce the problem of computing PRS into analyzing the distance between stochastic trajectories and their associated deterministic trajectories termed stochastic deviation. 
By developing a novel energy function called Averaged Moment Generating Function, we establish a tight high-probability bound on the stochastic deviation of stochastic systems. 
To the best of our knowledge, our bound is the tightest high-probability bound on stochastic deviation for general nonlinear systems. 
By combining this probabilistic bound on stochastic deviation with the contraction-based and interval-based reachability of deterministic systems, we provide tight estimates of PRS for stochastic systems.  
Our separation strategy and tight probabilistic bounds on stochastic deviation can transform many current methods/results in control theory and applications. 
They will also open new research directions in various fields, such as safety-critical control, estimation, uncertainty quantification, statistics, and machine learning.
Additionally, the AMGF leveraged in our theoretical analysis is a powerful mathematical tool, waiting for further exploitation in the future. 

\bibliographystyle{IEEEtran}
\bibliography{references,SJ}  

\appendix
\subsection{Proof of Lemma \ref{lemma: concentration} (Sub-Gaussian Norm Concentration)}\label{app: pf norm conc}
For every $\varepsilon\in(0,1)$, we can find a finite set $\mathcal{N}\subseteq\ball{1,0}$ such that for $\forall x_0\in\ball{1,0},~ \exists x\in\mathcal{N},~ \|x-x_0\|\leq\varepsilon$.
    Let $|\mathcal{N}|$ denote the number of elements in $\mathcal{N}$. By \cite[Exercise 4.4.2]{rigollet2023high}, there exists such an $\mathcal{N}$ that $|\mathcal{N}|\leq (1+2/\varepsilon)^n$ and for any vector $x\in\R^n$,
    \[
        \|x\|\leq \frac{1}{1-\varepsilon}\max_{\ell\in\mathcal{N}}\ell^{\top}x.
    \]
It follows that, for any $r>0$ and any sub-Gaussian vector $X\in\R^n$ with variance proxy $\sigma^2$, 
    \begin{equation}\label{eq: P|X|>r}
    \begin{split}
        &\prob{\|X\|\geq r}\leq\prob{\frac{1}{1-\varepsilon}\max_{\ell\in\mathcal{N}}\ell^{\top} X\geq r} \\
        \leq& \prob{\bigcup_{\ell\in\mathcal{N}}\frac{\ell^{\top}X}{1-\varepsilon}\geq r}.
    \end{split}
    \end{equation}
Since $\|\ell\|\leq1$ for $\ell\in\mathcal{N}$, we have
\begin{equation}\label{eq: l/l}
    \prob{\frac{\ell^{\top}X}{1-\varepsilon}\geq r,~ \ell\in\mathcal{N}}\leq \prob{\frac{\ell^{\top}X}{\|\ell\|(1-\varepsilon)}\geq r,~ \ell\in\mathcal{N}}.
\end{equation}
By the definition of sub-Gaussian vector, we know $\frac{\ell^{\top}X}{\|\ell\|}$ is sub-Gaussian with variance proxy $\sigma^2$ for any $\ell\in\mathcal{N}$. By Hoeffding's Inequality,
\begin{equation}\label{eq: l'x hoeff}
    \prob{\frac{\ell^{\top}X}{\|\ell\|(1-\varepsilon)}\geq r,~ \ell\in\mathcal{N}}\leq e^{-\frac{(1-\varepsilon)^2r^2}{2\sigma^2}}.
\end{equation}
Combining \eqref{eq: P|X|>r}-\eqref{eq: l'x hoeff} and taking union bound over $\ell\in\mathcal{N}$, we obtain 
\begin{equation}\label{eq: P|X|>r final}
    \begin{split}
        &\prob{\|X\|\geq r}\leq \prob{\bigcup_{\ell\in\mathcal{N}}\frac{\ell^{\top}X}{1-\varepsilon}\geq r} \\
        \leq& |\mathcal{N}|e^{-\frac{(1-\varepsilon)^2r^2}{2\sigma^2}} \leq
        (1+\frac{2}{\varepsilon})^ne^{-\frac{(1-\varepsilon)^2r^2}{2\sigma^2}}.
    \end{split}
    \end{equation}
To ensure a confidence level $\delta$, which means the right-hand side of \eqref{eq: P|X|>r final} $\leq\delta$, $r$ should satisfy
\begin{equation}
    r^2\geq \frac{2\sigma^2}{(1-\varepsilon)^2}(n\log(1+\frac{2}{\varepsilon})+\log\frac{1}{\delta}). 
\end{equation}
Then \eqref{eq: concentration norm} follows by taking the square root. This completes the proof.

\subsection{Proof of Lemma \ref{lemma: AMGF_1}} \label{pf: lemma AMGF_1}
\ref{p1:AMGF} For any $\lambda\in\R$ and $\eta_1,\eta_2\in\mS^{n-1}$, we have
\begin{equation}\label{eq: ineerp MGF}
    \mbE_{\ell\sim\mS^{n-1}}(e^{\lambda\innerp{\ell,\eta_1}})=\mbE_{\ell\sim\mS^{n-1}}e^{\lambda\innerp{\ell,\eta_2}}).
\end{equation}
It follows that, for any $x\in \real^n$ and $\eta\in\mS^{n-1}$, 
\begin{equation}\label{eq: r-invariant}
    \begin{split}
        \salf{x}&=\expectw{\ell\sim\mS^{n-1}}{e^{\lambda\innerp{\ell,x}}}
    \\ & = \expectw{\ell\sim\mS^{n-1}}{e^{\lambda\|x\|\innerp{\ell,\frac{x}{\|x\|}}}} \\ &= \expectw{\ell\sim\mS^{n-1}}{e^{\lambda\|x\|\innerp{\ell,\eta}}} \\ &= \salf{\|x\|\,\eta}.
    \end{split}
\end{equation}

\ref{p2:AMGF} By part \ref{p1:AMGF}, $\salf{x} = \salf{\|x\| \eta}$ for any $\eta\in\mS^{n-1}$. Taking the derivative of $\salf{x}$ over $\|x\|$ when $\|x\|\neq0$:
\begin{equation}\label{eq: cal dpsi/dx}
    \begin{split}
        &\frac{d\salf{x}}{d\|x\|}=\frac{d}{d\|x\|}\expectw{\ell\sim\mS^{n-1}}{e^{\lambda\|x\|\innerp{\ell,\eta}}} \\
        &=\expectw{\ell\sim\mS^{n-1}}{\lambda\innerp{\ell,\eta}e^{\lambda\|x\|\innerp{\ell,\eta}}} \\
        &=\frac{1}{\|x\|}\expectw{\ell\sim\mS^{n-1}}{\lambda\|x\|\innerp{\ell,\eta}e^{\lambda\|x\|\innerp{\ell,\eta}}}.
    \end{split}
\end{equation}
Set $y=e^{\lambda\|x\|\innerp{\ell,\eta}}$. Applying Jensen's Inequality over the convex function $y\log y$, we arrive at
\begin{equation*}
    \begin{split}
        &\expectw{\ell\sim\mS^{n-1}}{\lambda\|x\|\innerp{\ell,\eta}e^{\lambda\|x\|\innerp{\ell,\eta}}} \\
        \geq& \expectw{\ell\sim\mS^{n-1}}{\lambda\|x\|\innerp{\ell,\eta}} \, \expectw{\ell\sim\mS^{n-1}}{e^{\lambda\|x\|\innerp{\ell,\eta}}}=0.  
    \end{split}
\end{equation*}
Thus, $\frac{d\salf{x}}{d\|x\|}\geq0$ when $\|x\|\neq0$. When $\|x\|=0$, obviously $\salf{x}=1$ and $\frac{d\salf{x}}{d\|x\|}=0$. This completes the proof.

\subsection{Proof of Lemma \ref{lemma: exp.<>}} \label{pf: lemma exp.<>}
Let $\tau\mapsto x_{\tau}$ and $\tau\mapsto y_{\tau}$ be two trajectories of the system~\eqref{eq:associate-deterministic}. Since $\mu(D_x f(x,u,t))\le 0$, for every $x,u,t\in \real^n\times \mathcal{U}\times \real_{\ge 0}$, we get~\cite[Theorem 36]{AD-SJ-FB:20o} 
\begin{align*}
    \|x_{\tau}-y_{\tau}\| \le \|x_t-y_t\|,\mbox{ for all }\tau\ge t. 
\end{align*}
Using Lemma~\ref{lemma: AMGF_1}\ref{p2:AMGF}, for every $\tau\ge t$,  
\begin{align*}
    \mathbb{E}_{\ell\in \mS^{n-1}}(e^{\lambda \langle \ell, x_{\tau}-y_{\tau}\rangle}) \le \mathbb{E}_{\ell\in \mS^{n-1}}(e^{\lambda \langle \ell, x_t-y_t\rangle}).
\end{align*}
This implies that, for every $\tau\ge t$,  
\begin{align*}
    \mathbb{E}_{\ell\in \mS^{n-1}}\left(\tfrac{1}{\tau-t}\left(e^{\lambda \langle \ell, x_{\tau}-y_{\tau}\rangle}-e^{\lambda \langle \ell, x_{t}-y_{t}\rangle}\right)\right)\le 0.
\end{align*}
Taking the limit as $\tau-t\to 0^{+}$, we have
\begin{align*}
    \mathbb{E}_{\ell\in \mS^{n-1}}\left( e^{\lambda \langle \ell, x_t-y_t\rangle}\lambda\ell^{\top}(f(x_t,u_t,t)-f(y_t,u_t,t))\right)\le 0.
\end{align*}
The result follows by noting that $x_t,y_t\in\real^n$ and $u_t\in \mathcal{U}$ have been chosen arbitrarily. 

\subsection{Proof of Lemma \ref{lemma: salf to sg}} \label{pf: lemma salf2sg}
    Define random vector $\tilde{X}=QX$, where $Q\sim\mathbb{U}^n$ is a random unitary matrix. By Lemma \ref{lemma: AMGF_1}(i), we have that for any $\eta\in\mS^{n-1}$, 
        \begin{equation*}
        \begin{split}
            \salf{X}&=\expectw{\ell\sim\mS^{n-1}}{e^{\lambda\|X\|\innerp{\ell,\frac{X}{\|X\|}}}} =\expectw{\ell\sim\mS^{n-1}}{e^{\lambda\|X\|\innerp{\ell,\eta}}} \\
            &=\expectw{\ell\sim\mS^{n-1}}{e^{\lambda\innerp{\eta,\ell\|X\|}}} =\mbE_{Q\sim\mathbb{U}^n}\left(e^{\lambda\innerp{\eta,QX}}\right),
        \end{split}
        \end{equation*}
    where the last ``$=$'' uses the fact that $Q\ell\in\mS^{n-1}$ for any $\ell\in\mS^{n-1}$. By \eqref{eq: lemma salf_2}, we obtain
    \begin{equation}\label{eq: Qx MGF half}
        \begin{split}
    &\mbE_{\tX}\left(e^{\lambda\innerp{\eta,\tX}}\right)=\mbE_X\mbE_Q\left(e^{\lambda\innerp{\eta,QX}}\right) \\
        = &\mbE_X\left(\salf{X}\right)\leq e^{\frac{\lambda^2\sigma^2}{2}},\quad\forall \lambda\in\R,~\forall \eta\in\mS^{n-1}.
        \end{split}
    \end{equation}
    Therefore, $\tilde{X}$ is sub-Gaussian with variance proxy $\sigma^2$. By Lemma \ref{lemma: concentration}, $\tilde{X}$ satisfies \eqref{eq: concentration norm}. 
    
    Finally, since $\|X\|=\|QX\|=\|\tilde{X}\|$ for any $Q\in\mathbb{U}^n$, we conclude that $X$ also satisfies \eqref{eq: concentration norm}. This completes the proof.

\end{document}